%% file: main.tex
\documentclass[twocolumn,dvipsname]{aastex702}

\usepackage[version=4]{mhchem}
\usepackage{siunitx}
\usepackage{xspace}
\usepackage{amsmath}
\usepackage{wasysym}
\usepackage{booktabs}
\usepackage{colortbl}
\definecolor{table_header}{RGB}{232,239,245}
\definecolor{table_accent}{RGB}{245,248,250}

\let\oldAA\AA
\renewcommand{\AA}{{\text{\normalfont\oldAA}}\xspace}

\input{define.tex} 

\begin{document}

\title{The 1.6~$\mu\mathrm{m}$ \hminus{} bump as a diagnostic of TP-AGB contributions and recent quenching. I. Insights from stellar population models}

\author[orcid=0009-0000-4701-4934]{Pengjun Lu}
\affiliation{Department of Astronomy, Tsinghua University, Beijing 100084, China}
\email{lupj23@mails.tsinghua.edu.cn}

\author[0000-0003-1385-7591]{Song Huang}
\affiliation{Department of Astronomy, Tsinghua University, Beijing 100084, China}
\email{shuang@tsinghua.edu.cn}

\author[0009-0003-4133-0292]{Sijia Cai} 
\affiliation{Department of Astronomy, Tsinghua University, Beijing 100084, China}
\email{caisj23@mails.tsinghua.edu.cn}

\author[orcid=0000-0003-2388-8172,sname='Francesco D'Eugenio']{Francesco D'Eugenio}
\affiliation{Kavli Institute for Cosmology, University of Cambridge, Madingley Road, Cambridge, CB3 0HA, UK}
\affiliation{Cavendish Laboratory, University of Cambridge, 19 JJ Thomson Avenue, Cambridge, CB3 0HE, UK}
 \email{francesco.deugenio@gmail.com}

\author[orcid=0000-0002-4267-9344]{Meng Gu}
\affiliation{Department of Astronomy, Tsinghua University, Beijing 100084, China}
\affiliation{Hong Kong Institute for Astronomy \& Astrophysics, Pokfulam Road, The University of Hong Kong, Pok Fu Lam, Hong Kong}
\email{menggu@tsinghua.edu.cn}

\author[orcid=0009-0006-1447-2625]{Li Shen}
\affiliation{Department of Astronomy, Tsinghua University, Beijing 100084, China}
\email{l-shen26@mails.tsinghua.edu.cn}

\author[orcid=0000-0002-8711-8970]{Cheng Li}
\affiliation{Department of Astronomy, Tsinghua University, Beijing 100084, China}
\email{cli2015@tsinghua.edu.cn}

\correspondingauthor{{Song Huang}}
\email{shuang@tsinghua.edu.cn}

\begin{abstract}
    The 1.6~\si{\micro\metre} \hminus{} bump arises from a minimum in the opacity of the negative hydrogen ion in cool stellar atmospheres. Although it is recognized as a common feature in galaxy spectra, its potential for studying stellar populations and recent quenching remains underexplored. Its enhancement at intermediate ages is strong in some stellar population models with substantial thermally pulsing asymptotic giant branch (TP-AGB) contributions but weak or absent in others. This variation makes the feature both a potential age diagnostic and a source of model uncertainty. We use models based on the X-shooter Spectral Library (XSL) and Flexible Stellar Population Synthesis (FSPS) to examine what drives this behavior and what it can reveal about quenching. Controlled FSPS experiments show that the \hminus{} bump depends on both the amount of TP-AGB light and the stellar template used to represent it. The adopted XSL and TP-AGB-on FSPS models predict a transient enhancement of the \hminus{} bump following rapid quenching, while optical spectral features $\mathrm{D}4000$ and H$\delta$ produce monotonic responses. Combining these diagnostics can therefore provide additional information about recent quenching, although the interpretation depends on the population model and star-formation history. The \hminus{} bump -- $\mathrm{D}4000$ plane shows that models including TP-AGB contributions can describe the observed JWST NIRSpec data reasonably well. These results motivate further use of the \hminus{} bump to test TP-AGB prescriptions and investigate quenching with current and forthcoming observations.
\end{abstract}


\section{Introduction} 
    \label{sec:intro}

    Understanding how rapidly galaxies cease forming stars can help distinguish the processes responsible for quenching. Environmental effects, feedback from active galactic nuclei (AGN) and stars, and the depletion or removal of cold gas may all contribute. Their relative importance remains uncertain, as illustrated by the challenges of explaining both massive quiescent galaxies at high redshift \citep[e.g.,][]{Carnall-2023, Valentino-2023, Carnall-2024} and the local passive population \citep[e.g.,][]{Wild-2016, Pawlik-2018}.

    Post-starburst galaxies provide a view of this transition shortly after substantial star formation has ceased \citep[e.g.,][]{Chen-2019, Leung-2024}. Classical E+A selection combines strong Balmer absorption with weak or absent nebular emission \citep{Dressler-1983}. More generally, H$\delta$ and $\mathrm{D}4000$ trace different aspects of recent star-formation history and are widely used to identify post-starburst and rejuvenating populations \citep[e.g.,][]{Wu-2018, Zhang-2023}. In this context, near-infrared (NIR) features may add information through a different stellar component. In some models, thermally pulsing asymptotic giant branch (TP-AGB) stars contribute substantially to the NIR light at stellar ages of approximately 0.5--2~Gyr \citep[e.g.,][]{Maraston-2005a, Noel-2013, Verro-2022a}. As young stars fade after quenching, features associated with these intermediate-age populations may become more prominent, with the response depending on the preceding history and the quenching timescale.

    TP-AGB stars can contribute substantially to the NIR luminosity of intermediate-age stellar populations. In this work, we focus on the broad NIR continuum feature around 1.6~\si{\micro\metre}, known as the \hminus{} bump. Its origin is a minimum in the continuous opacity of the negative hydrogen ion, \ce{H-}, in cool stellar atmospheres \citep{John-1988, Sawicki-2002}. Near this wavelength, bound-free absorption falls toward its long-wavelength threshold, while free-free absorption increases toward longer wavelengths. The resulting opacity minimum allows radiation to escape from deeper, hotter atmospheric layers, producing a local enhancement in the emergent spectrum. Molecular absorption, including \ce{H2O}, further shapes the surrounding spectrum and affects the apparent contrast of the \hminus{} bump. The feature is present in cool red giants and in both oxygen-rich and carbon-rich TP-AGB stars \citep{Verro-2022a, Lu-2025, Lu-2026}. Its strength in integrated galaxy light therefore depends on the relative contributions and spectral properties of these stars. In populations with substantial TP-AGB contribution, this can produce a pronounced intermediate-age signature, becoming a potential intermediate-age spectral tracer.

    As a stellar continuum feature, the 1.6~\si{\micro\metre} bump has also been explored as a photometric redshift indicator and used for infrared color selection because its characteristic wavelength provides a reference in galaxy spectral energy distributions \citep{Sawicki-2002, Sorba-2010}. More recently, work using simulated SPHEREx data has identified the 1.6~\si{\micro\metre} region as particularly useful for stellar mass estimation \citep{Lee-2025}. The underlying opacity physics also has applications beyond integrated stellar populations. \citet{Liu-2026} predict a related 1.6~\si{\micro\metre} ``kink'' in optically thick atmosphere models for little red dots and show that its strength depends on photospheric density. In that setting, the feature probes a proposed gaseous atmosphere. This connection extends the relevance of \ce{H-} opacity to a different class of extragalactic sources. However, beyond intermediate-age populations, using the \hminus{} bump to trace stellar ages and recent quenching remains relatively underexplored. 

    Studying quenching signatures using this feature should take into account the uncertainties in TP-AGB evolution models. Observational studies have reached differing conclusions about the NIR contribution of TP-AGB stars \citep{Kriek-2010, Melbourne-2012, Zibetti-2013, Riffel-2015, Bevacqua-2025, Lu-2025, Bevacqua-2025, Lu-2026}. These comparisons depend on both population properties and the spectral libraries used \citep{Baldwin-2018}. The model-dependent intermediate-age enhancement of the \hminus{} bump therefore motivates testing both TP-AGB light fractions and assigned spectra. Moreover, its broad profile makes such tests accessible at low spectral resolution.

    In this first paper, we ask whether the \hminus{} bump can test the contribution and spectral treatment of TP-AGB stars, and whether combining it with optical age indicators adds information about recent quenching. X-shooter Spectral Library (XSL) models, with empirical TP-AGB spectra and PARSEC/COLIBRI evolution, predict its time dependence. Controlled FSPS experiments allow us to vary the TP-AGB contribution and quantify the additional information. In Section~\ref{sec:data}, we describe the models and measurements; Section~\ref{sec:results} presents the quenching predictions and mock tests. Section~\ref{sec:discuss} discusses model limitations, compares the predictions with JWST observations, and considers observational prospects. Section~\ref{sec:conclude} summarizes our conclusions.

\section{Models and Analysis} 
    \label{sec:data}

\subsection{The X-shooter Spectral Library}
    \label{ssec:xshooter}

    The X-shooter Spectral Library (XSL) is an empirical library for stellar population synthesis\footnote{\url{http://xsl.u-strasbg.fr}}. Its spectra cover 350 to 2480 nm at $R\sim10{,}000$, allowing optical and near-infrared features to be studied within the same library. XSL includes cool evolved stars that can contribute strongly to the near-infrared light of intermediate-age populations. Its samples of oxygen-rich static giants, oxygen-rich TP-AGB stars, and carbon-rich TP-AGB stars allow these phases to be treated separately. In the SSP framework of \citet{Verro-2022a}, 44 static O-rich giants, 39 O-rich TP-AGB stars, and 26 carbon-rich TP-AGB spectra are incorporated into dedicated spectral sequences. These sequences provide an empirical basis for studying TP-AGB-sensitive near-infrared features, including the \hminus{} bump and molecular absorption bands.

    We first examine representative XSL stellar spectra from Data Release 3 \citep{Verro-2022b}, obtained with the X-shooter three-arm spectrograph on ESO's VLT \citep{Vernet-2011}. From the published 830 spectra of 683 stars, we randomly selected spectra of A-type, G-type, RGB, C-rich, and O-rich TP-AGB stars, aiming to cover a wide range of effective temperatures. In the inset, we divide the spectra around the \hminus{} bump by the linear pseudo-continuum to illustrate the feature strength.

\begin{figure}[htbp]
    \centering
    \includegraphics[width=\columnwidth]{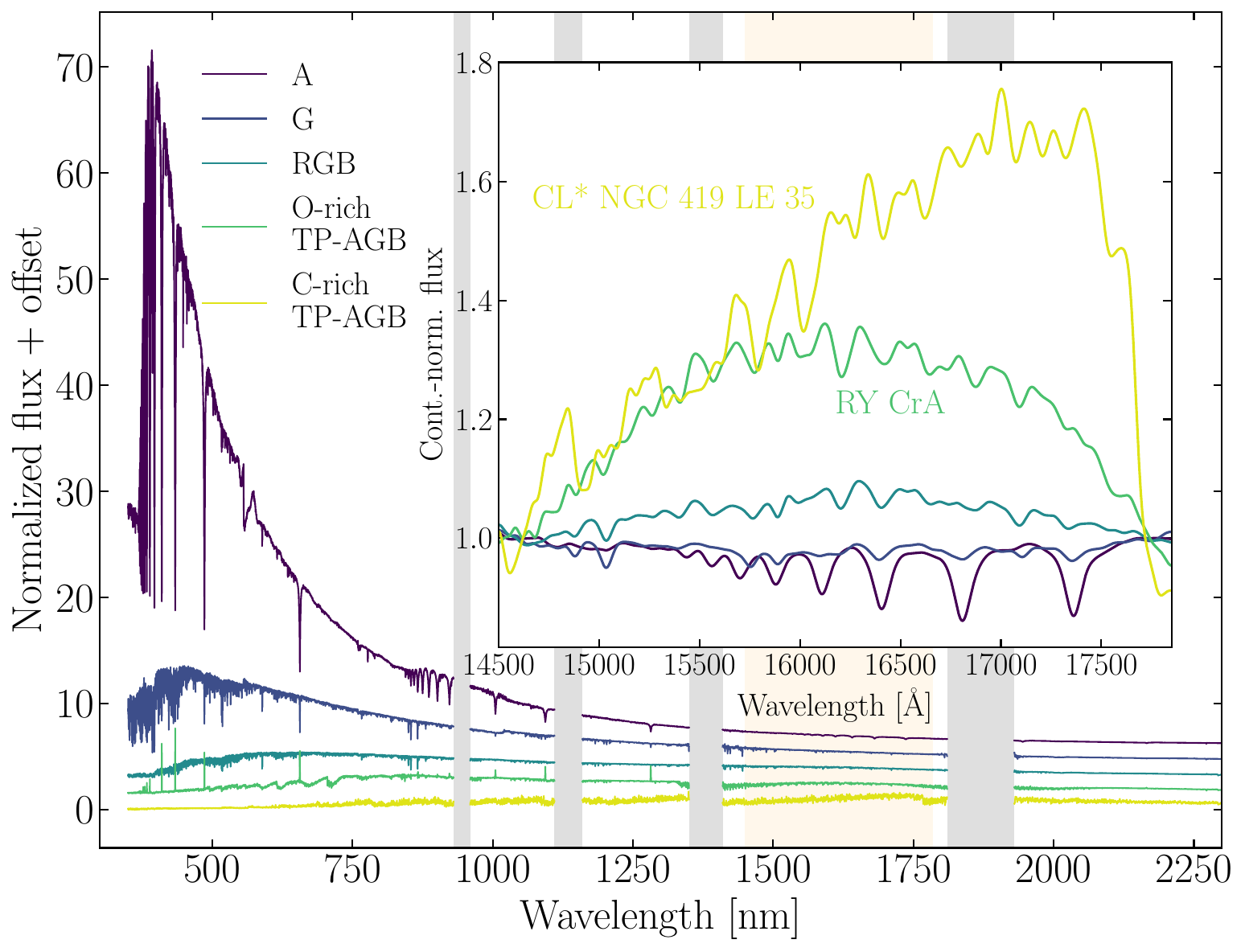}
    \caption{
        Spectra of five stellar types (A, G, RGB, O-rich TP-AGB, and C-rich TP-AGB), observed with the X-shooter spectrograph; line colors are shown in the legend. The inset panel shows the zoomed-in spectrum around 1.6 \si{\micro\metre}, normalized to a linear pseudo-continuum fitted between the blue and red continuum windows. As illustrated, the O-rich TP-AGB star is RY CrA, and the C-rich TP-AGB star is $\rm{CL}^*$NGC 419 LE 35.}
\label{stellar}
\end{figure}

    As shown in Figure \ref{stellar}, the theoretical temperature dependence of \hminus{} opacity \citep{John-1988} is consistent with the trend seen in the observed stellar spectra.  Among the illustrated giant-star spectra, the cooler stars show a more prominent \hminus{} bump. The RGB spectrum shows an excess above the pseudo-continuum. Moreover, the bump is much stronger in the observed TP-AGB spectra as illustrated. This temperature-sensitive behavior, as predicted by theories, motivates us to analyze its performance on stellar population ages.

\subsection{Simple Stellar Populations}
    \label{ssec:ssp}

    We adopt the archived XSL SSP models described by \citet{Verro-2022a}, using the PARSEC/COLIBRI isochrones and the Kroupa initial mass function (IMF) \citep{Kroupa-2001}. The COLIBRI models implement a detailed treatment of the TP-AGB phase, following stellar evolution from the first thermal pulse through envelope evolution, mass loss, and changes in surface chemistry, including the transition to carbon-rich stars. The SSP spectra cover 350--2480 nm at $R\sim10{,}000$. We use models with ages $\gtrsim50$ Myr and [Fe/H] from approximately $-2.2$ to $+0.2$. 

    For comparison with observations and other SSP models, we smooth the SSP spectra with FFT-based convolution implemented in \texttt{sedpy}. Our baseline calculations use Gaussian velocity broadening at resolution $\sigma = 300~\rm km/s$; we also downgrade spectra to $R=100$ and $R=50$ for the separate lower-resolution comparisons as described below. We sample the smoothed spectra at 3 pixels per resolution element. For index comparison among models, we also adopt E-MILES \citep{Rock-2016} SSPs with Padova00 isochrone \citep{girardi2000evolutionary}, C3K SSPs \citep{Conroy-2018} with MIST isochrone \citep{Choi-2016, dotter2016mesa} and M13 SSPs \citep{Noel-2013, newman_2025_13744975} \footnote{``maraston13\_kroupa\_cloudy-c23.01-sps.hdf5'' from: \url{https://zenodo.org/records/13744975}}. Although choosing between the Kroupa and Salpeter \citep{Salpeter-1955} IMFs has negligible impact on the measured \hminus{} bump strength in these grids, we choose the Kroupa IMF \citep{Kroupa-2001} for consistency. 

\subsection{Composite Stellar Populations}
    \label{ssec:csp}

    To follow the \hminus{} bump during quenching, we construct composite stellar populations (CSPs) with a delayed-exponential star-formation history (SFH) followed by exponential decline. Up to an overall normalization,
    \begin{equation*}
        \mathrm{SFR}(t)=\begin{cases}
        \frac{t}{\tau} e^{-t/\tau}, & t < t_q \\
        \frac{t_q}{\tau} e^{-t_q/\tau}e^{-(t-t_q)/\tau_q}, & t \ge t_q
    \end{cases}
    \end{equation*}

    Here, $t$ is time since the start of star formation, $t_q$ marks the quenching onset, and $\tau_q$ is the subsequent e-folding timescale. We set $\tau=t_q$, so the SFR peaks at quenching onset. At observation time $t_{\rm obs}$, the elapsed quenching time is $t_{\rm obs}-t_q$, whereas stars formed at $t'$ have age $t_{\rm obs}-t'$.

    We construct CSP spectra according to the SFH using custom code for XSL and the internal SFH routines of FSPS for the FSPS-based populations. Appendix~\ref{sec:appendixA} describes the construction and the consistency check between the two implementations.

\subsection{FSPS mock populations}
    \label{ssec:fsps_mocks}

    For the controlled tests in Section~\ref{sec:index}, we use \texttt{FSPS} v4.0 \citep{Conroy-2009, Conroy-2010} with MIST isochrones, the high-resolution C3K v2.3 library, and a Chabrier IMF \citep{Chabrier-2003}. ``TP-AGB off'' has \texttt{agb}=0, removing TP-AGB light but retaining other evolved stars. ``TP-AGB on'' has \texttt{agb}=2 and \texttt{use\_lw\_tpagb}=1, assigning the empirical O-rich spectra of \citet{Lancon-2002} to cool TP-AGB stars at twice their fiducial weight. We hold the isochrones, IMF, and other stellar components fixed. These deliberately contrasting configurations test sensitivity to the TP-AGB treatment.

    During the CSP construction, we keep the metallicity fixed throughout each FSPS history. We omit nebular emission and interstellar dust attenuation but retain the default circumstellar AGB dust treatment. We summarize CSP generation and spectral broadening in Appendix~\ref{sec:appendixA}. For the experiments in Section~\ref{sec:index}, $\mathrm{D}4000$ and H$\delta_{\rm A}$ are measured at a target Gaussian velocity dispersion of $300~\mathrm{km\,s^{-1}}$. The \hminus{} bump is measured at $R=100$.

    We sampled CSP populations using the parameter distributions in Table~\ref{param-distribution}. The reference FSPS ensemble contains 2000 independent draws of the delayed-$\tau$ histories, followed by an exponential decline as defined in Section~\ref{ssec:csp}, with $\tau=t_q$. Each history is evaluated every 50~Myr from $t_{\rm obs}=0.05$ to 13~Gyr, and we retain $t_{\rm obs}\geq1$~Gyr for the population analysis. This gives 482\,000 sampled epochs per configuration, with the same SFHs and metallicities in both configurations. Instead of mocking the galaxy populations across redshifts, we give each epoch equal weight in the sampling tests.

\begin{table}[htbp]
    \centering
    \caption{Parameter sampling for the reference FSPS mock ensemble.}
    \label{param-distribution}
    \renewcommand{\arraystretch}{1.25}
    \setlength{\tabcolsep}{4pt}
    \begin{tabular}{llr}
    \toprule
    \rowcolor{table_header}[\tabcolsep][\tabcolsep]
    \textbf{Parameter} & \textbf{Distribution} & \textbf{Range} \\
    \midrule
    $t_q$ & Uniform & 1--5.9 Gyr \\
    $\tau_q$ & Log-uniform & 0.1--3 Gyr \\
    $\log_{10}(Z/Z_\odot)$ & Truncated normal & $-0.5$ to $+0.2$ \\
    $t_{\rm obs}$ & 50 Myr spacing & 1--13 Gyr \\
    \bottomrule
    \end{tabular}
    \tablecomments{
        The parent metallicity distribution has mean zero and standard deviation 0.2 dex. The time grid lists retained epochs for each of 2000 histories. Overall mass normalization cancels from the indices. We derived sSFRs using the surviving stellar mass at each epoch.}
\end{table}

    We label each epoch using its recent and previous star formation, following \citet{Zhang-2023}. Let $s_{\rm recent}$ and $s_{\rm previous}$ denote the SFRs averaged over lookback intervals of 0--100~Myr and 100--1000~Myr, respectively, each divided by the current mass in living stars and remnants. We define $\mathcal{R}=s_{\rm recent}/s_{\rm previous}$. The rapid-quenching class satisfies $s_{\rm previous}>10^{-10}~\mathrm{yr^{-1}}$ and $\mathcal{R}<0.1$: its recent average SFR is less than one tenth of the preceding average. Epochs with $s_{\rm previous}<10^{-10}~\mathrm{yr^{-1}}$ and $s_{\rm recent}<10^{-11}~\mathrm{yr^{-1}}$ are quiescent. Among the remaining epochs, those with $\mathcal{R}>1$ are labeled star-forming and all others transitional. The rapid-quenching class contains 5907 epochs, or 1.23\% of this ensemble, all from histories with $\tau_q<0.27$~Gyr. We note that this fraction depends on the adopted SFH and time sampling. Parameter distributions should be treated more carefully to account for physical duty cycles.

\subsection{Measurements of H\texorpdfstring{$\rm \delta$}{delta} and \texorpdfstring{$\mathrm{D}4000$}{D4000}}
    \label{sec:features}

    To quantify the age sensitivity of the \hminus{} bump, we compare it with the widely used stellar age indicators H$\rm \delta$ and $\mathrm{D}4000$, across SSP grids spanning age and metallicity. We use the wide-band definition of $\mathrm{D}4000$ \citep{Bruzual-1983} and the Lick-style H$\delta$ absorption index \citep{1997ApJS..111..377W, 2003MNRAS.339..897T}:

    \begin{equation*}
        \mathrm{D}4000 = \frac{\lambda_{b_2}-\lambda_{b_1}}{\lambda_{r_2}-\lambda_{r_1}}\frac{\int^{\lambda_{r_2}}_{\lambda_{r_1}}\lambda^2 F(\lambda)d\lambda}{\int^{\lambda_{b_2}}_{\lambda_{b_1}}\lambda^2 F(\lambda)d\lambda}
    \end{equation*}

    \begin{equation*}
        I_{\rm H\delta}(\AA) = \int^{\lambda_{c_2}}_{\lambda_{c_1}}\left(1-\frac{F(\lambda)}{F_\mathrm{c}(\lambda)}\right)d\lambda
    \end{equation*}

    where \(F(\lambda)\) denotes the flux density per unit wavelength, \(F_\lambda\). Same as \texttt{pyphot} \citep{Fouesneau-2026}\footnote{Code: \url{https://github.com/mfouesneau/pyphot}; index implementation: \url{https://mfouesneau.github.io/pyphot/licks.html}.}, We measure H$\delta$ and the \hminus{} bump with using a first-degree polynomial (\texttt{degree=1}) for the pseudo-continuum. A single straight line is fitted jointly to all spectral flux-density samples in the blue and red pseudo-continuum windows, and this fitted line defines \(F_\mathrm{c}(\lambda)\) for both model and observed spectra. The adopted H$\delta_{\rm A}$ feature band is 4083.50--4122.50~\AA, with all band limits listed in Table~\ref{indice-window}. The optical band limits are converted from air to vacuum wavelengths to match the FSPS spectra. $\mathrm{D}4000$ is the ratio of wavelength-averaged $F_\nu$ in its red and blue bands, equivalent to the expression above.

\subsection{Measurements of the \texorpdfstring{$\rm{H}^{-}$} bump}
\label{sec:hbump}
    
    There is no standard spectral index for the \hminus{} bump. We adopt the magnitude form used by \citet{Verro-2022a} and \citet{Lu-2026}:
    
    \begin{equation*}
        I_{\rm mag}(\mathrm{mag}) = -2.5\log_{10}\left(\frac{1}{\lambda_{c_2}-\lambda_{c_1}}\int^{\lambda_{c_2}}_{\lambda_{c_1}}\frac{F(\lambda)}{F_c(\lambda)}d\lambda\right)
    \end{equation*}

    We use the same straight-line pseudo-continuum fit described above and report the index in magnitudes. The blue pseudo-continuum, central feature, and red pseudo-continuum windows are specified in rest-frame vacuum wavelengths.

    This definition follows the absorption-index sign convention: a mean flux excess above the pseudo-continuum gives \(I_{\rm mag}<0\), and a stronger \hminus{} bump corresponds to a more negative index. We report \(I_{\rm mag}\), in magnitudes, for the model and observational measurements. For the window-selection objective, we use the absolute value of \(-I_{\rm mag}\), \(H=-I_{\rm mag}\), so that larger \(H\) corresponds to a stronger \hminus{} bump.
    
    The narrow windows of \citet{Verro-2022a} suit high-resolution stellar spectra ($R \sim 10000$) but often contain too few samples for reliable integration in low-resolution galaxy spectra. To reduce this sensitivity to spectral resolution, \citet{Lu-2026} adopted broader wavelength windows. Here we explore a flexible framework for selecting the blue pseudo-continuum, feature, and red pseudo-continuum windows of the \hminus{} bump index to highlight the age sensitivity. Candidate definitions are ranked by their sensitivity to stellar age relative to their dependence on other population parameters and spectral resolution. We demonstrate the method with the adopted XSL SSP grids, seeking an illustrative index that reveals age-sensitive information in the intermediate-age region. The preferred definition will depend on the scientific objective, stellar population models, and observational conditions.

    For this demonstration, we evaluate the auxiliary quantity \(H(i,R,Z,t)\) defined above over IMF, resolution, metallicity, and age. We adopt an objective function that rewards the summed contrast of the \(1\,{\rm Gyr}\) population against the \(0.1\) and \(10\,{\rm Gyr}\) populations, while penalizing sensitivity to metallicity, IMF, and spectral resolution:
        \[
        \mathcal{L}
        =
        \log S_t
        -\log S_Z
        -\log S_{\rm IMF}
        -\log S_R .
        \]
    Here \(S_t\) measures the age contrast around \(1\,{\rm Gyr}\). \(S_Z\), \(S_{\rm IMF}\), and \(S_R\) quantify variation with metallicity, IMF, and resolution, respectively, averaged over ages of \(0.5\)--\(2\,{\rm Gyr}\) and the remaining parameters (see Table \ref{tab:hbump_sensitivity_terms}). The objective ranks candidate definitions according to this adopted balance of sensitivities.

    \begin{table}[ht]
    \centering
    \caption{Definitions of the \hminus{} bump sensitivity terms, where
    $\delta H_{1-0.1\,{\rm Gyr}}
    =
    H(i,R,Z,1\,{\rm Gyr}) - H(i,R,Z,0.1\,{\rm Gyr})$, and
    $\delta H_{1-10\,{\rm Gyr}}
    =
    H(i,R,Z,1\,{\rm Gyr}) - H(i,R,Z,10\,{\rm Gyr})$.}
    \label{tab:hbump_sensitivity_terms}
    \renewcommand{\arraystretch}{1.3}
    \setlength{\tabcolsep}{5pt}
    \begin{tabular}{cc}
    \toprule
    \rowcolor{table_header}[\tabcolsep][\tabcolsep]
    \textbf{Sensitivity term} & \textbf{Definition} \\
    \midrule
    \addlinespace[0.35em]
    \(S_t\) &
    \(\displaystyle
    \left\langle
    \delta H_{1-0.1\,{\rm Gyr}} + \delta H_{1-10\,{\rm Gyr}}
    \right\rangle_{i,R,Z}
    \) \\[0.8em]
    
    \(S_Z\) &
    \(\displaystyle
    \left\langle
    {\rm std}_{Z}\!\left[H(i,R,Z,t)\right]
    \right\rangle_{i,R,t\in[0.5,2]\,{\rm Gyr}}
    \) \\[0.8em]
    
    \(S_{\rm IMF}\) &
    \(\displaystyle
    \left\langle
    {\rm std}_{i}\!\left[H(i,R,Z,t)\right]
    \right\rangle_{R,Z,t\in[0.5,2]\,{\rm Gyr}}
    \) \\[0.8em]
    
    \(S_R\) &
    \(\displaystyle
    \left\langle
    {\rm std}_{R}\!\left[H(i,R,Z,t)\right]
    \right\rangle_{i,Z,t\in[0.5,2]\,{\rm Gyr}}
    \) \\
    \bottomrule
    \end{tabular}
    \end{table}

    We adopt a two-stage search: first sampling valid random window triplets within the configured wavelength bounds, and then refining the highest-scoring candidates with Gaussian perturbations. The highest-scoring candidate \hminus{} bump definition among the windows evaluated in this demonstration is

    \begin{equation*}
        \begin{aligned}
        \lambda_{\rm blue}=1.494 -1.539\,\mu{\rm m},\\
        \lambda_{\rm feat}=1.570 -1.734\,\mu{\rm m},\\
        \lambda_{\rm red}=1.746 -1.791\,\mu{\rm m},          
        \end{aligned}
    \end{equation*}

    with \(\mathcal{L}=10.27\), compared with scores of \(8.25\) and \(8.00\) for the Lu2026 and Verro2022 definitions under the same objective and model grid (see Appendix \ref{sec:appendixB} for details). We denote this definition as Hbump\_opt, where ``opt'' identifies the highest-scoring candidate evaluated in our two-stage search using the adopted XSL grid, objective function, and window constraints. Table~\ref{indice-window} summarizes the wavelength windows for H$\rm \delta$, $\mathrm{D}4000$, and the \hminus{} bump definitions.

    \begin{table*}[t]
    \centering
    \caption{Rest-frame wavelength bands of the spectral indices.}
    \label{indice-window}
    \renewcommand{\arraystretch}{1.25}
    \setlength{\tabcolsep}{5pt}
    \begin{tabular}{lc*{6}{r}}
    \toprule
    \rowcolor{table_header}[\tabcolsep][\tabcolsep]
    \textbf{Index} & \textbf{Unit} & \multicolumn{2}{c}{\textbf{Blue band}} & \multicolumn{2}{c}{\textbf{Feature band}} & \multicolumn{2}{c}{\textbf{Red band}} \\
    \rowcolor{table_header}[\tabcolsep][\tabcolsep]
    & & \multicolumn{1}{c}{Lower} & \multicolumn{1}{c}{Upper} & \multicolumn{1}{c}{Lower} & \multicolumn{1}{c}{Upper} & \multicolumn{1}{c}{Lower} & \multicolumn{1}{c}{Upper} \\
    \midrule
    $\mathrm{D}4000$ & \AA & 3750.00\phantom{0} & 3950.00\phantom{0} & \multicolumn{2}{c}{\textemdash} & 4050.00\phantom{0} & 4250.00\phantom{0} \\
    H$\delta_{\rm A}$ & \AA & 4041.60\phantom{0} & 4079.75\phantom{0} & 4083.50\phantom{0} & 4122.50\phantom{0} & 4128.50\phantom{0} & 4161.00\phantom{0} \\
    \midrule
    Hbump\_Verro22 & $\mu\mathrm{m}$ & 1.450 & 1.470 & 1.610 & 1.670 & 1.765 & 1.785 \\
    Hbump\_Lu26 & $\mu\mathrm{m}$ & 1.400 & 1.500 & 1.500 & 1.800 & 1.800 & 2.000 \\
    \rowcolor{table_accent}[\tabcolsep][\tabcolsep]
    \textbf{Hbump\_opt} & $\mu\mathrm{m}$ & 1.494 & 1.539 & 1.570 & 1.734 & 1.746 & 1.791 \\
    \bottomrule
    \end{tabular}
    \tablecomments{The $\mathrm{D}4000$ and H$\delta_{\rm A}$ limits are listed in air wavelengths and converted to vacuum for measurements on the FSPS spectra. The \hminus{} bump limits are in vacuum wavelengths. The blue and red bands define the pseudo-continuum for H$\delta_{\rm A}$ and the \hminus{} bump; $\mathrm{D}4000$ uses their flux ratio and has no central feature band. The three \hminus{} bump definitions are Verro22~\citep{Verro-2022a}, Lu26~\citep{Lu-2026}, and Hbump\_opt from this paper. We adopt Hbump\_opt for the following results, whose selection and the common pseudo-continuum fitting procedure are described in Section~\ref{sec:hbump}.}
    \end{table*}

\section{Results}
    \label{sec:results}

\subsection{Model-dependent age sensitivity of the \texorpdfstring{$\rm{H}^{-}$} bump}
    \label{ssec:grids}

\begin{figure*}[htbp]            
    \includegraphics[width=1\textwidth]{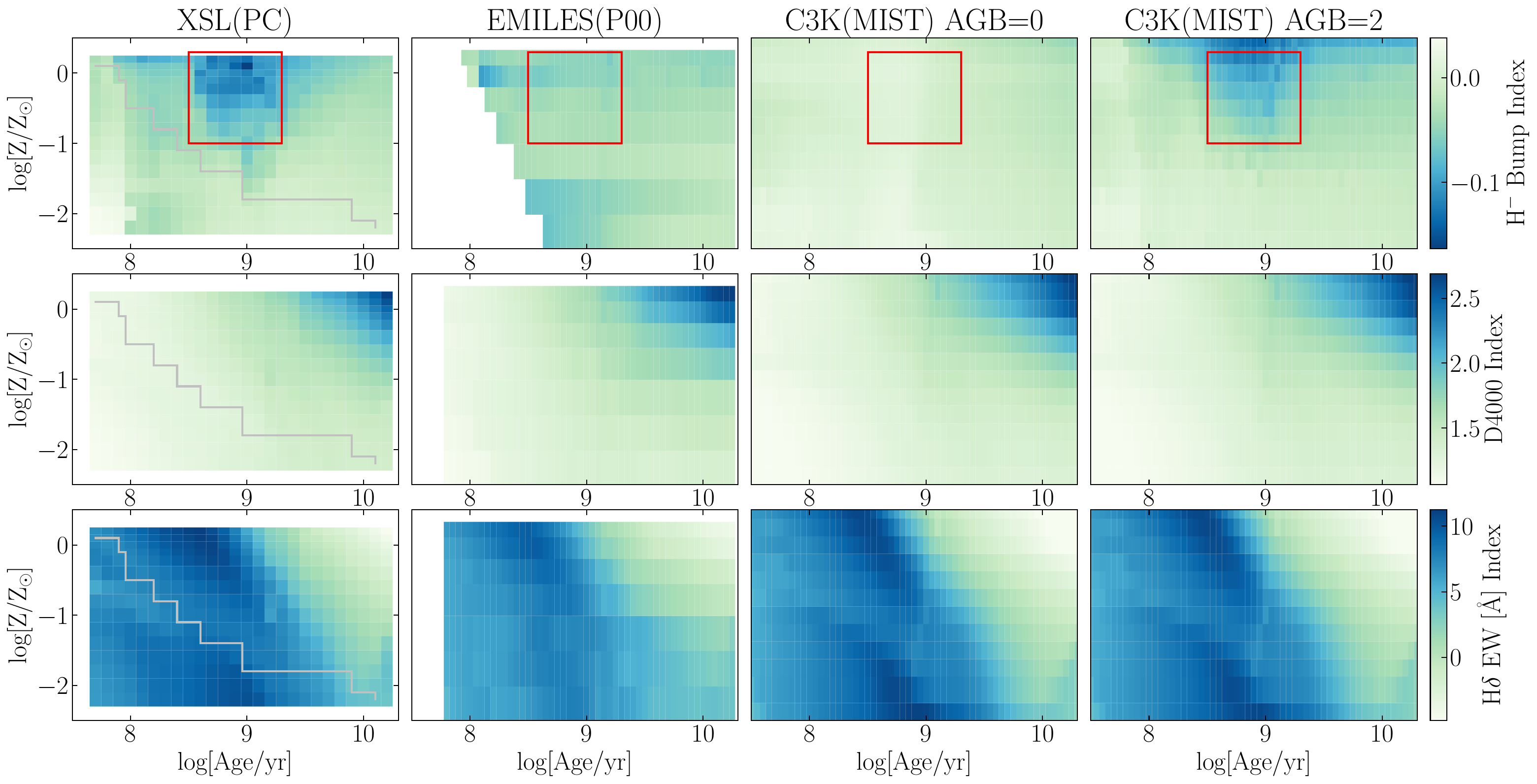}
    \caption{Spectral feature indices of simple stellar population (SSP) models in the plane of age and metallicity, measured at $\sigma = 300~\rm km/s$. From top to bottom, the rows show the \hminus{} bump, \(\mathrm{D}4000\), and H\(\delta\) indices, respectively, as a function of stellar population age and metallicity. The first two columns show XSL combined with the PARSEC/COLIBRI (PC) isochrones and E-MILES with the Padova00 isochrones. The last two columns show C3K+MIST SSPs generated with \texttt{FSPS}, with the LW02 O-rich TP-AGB templates enabled and the weighting parameter set to \texttt{agb}=0 and 2, respectively; the remaining settings are held fixed. These cases suppress and enhance the TP-AGB contribution relative to the default weight of unity and illustrate the sensitivity of the indices to this contribution. The gray boundaries in the first column delineate the ``safe zone'' of the XSL+PC model, with the upper-right region enclosed by these boundaries representing the reliable parameter space. The \hminus{} bump color scale represents \(I_{\rm mag}\) in magnitudes. The red rectangles highlight the intermediate-age region where the \hminus{} bump is strongest with substantial TP-AGB contributions, corresponding to the most negative index values.}
\label{SSP}
\end{figure*}

\begin{figure}[htbp]            
    \includegraphics[width=\columnwidth]{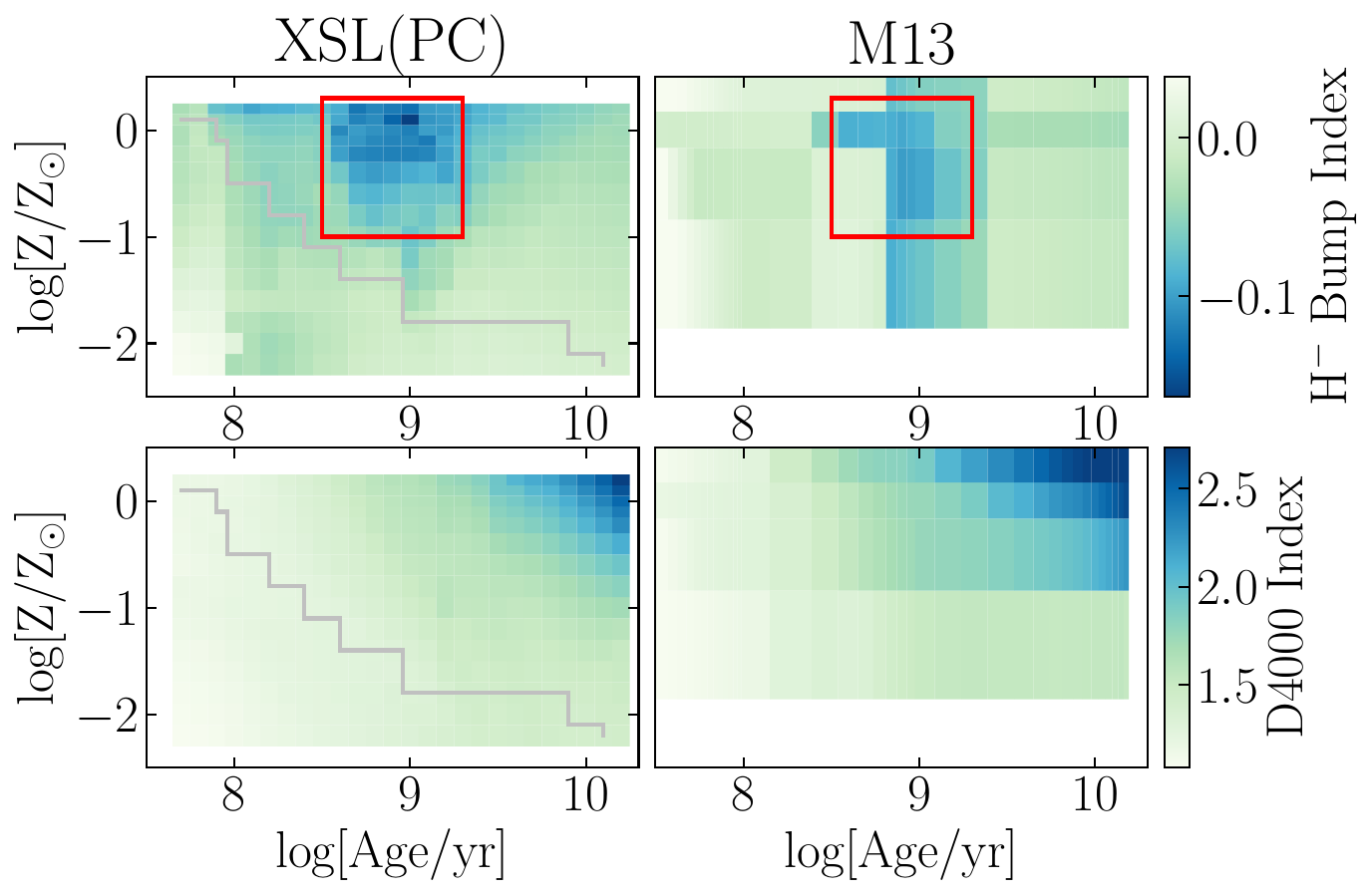}
    \caption{Similar to Figure \ref{SSP}, the rows are the \hminus{} bump index \(I_{\rm mag}\) and $\mathrm{D}4000$, and the columns are the XSL+PC model and the M13 model. All models are measured at $R=100$.}
\label{M13-SSP}
\end{figure}

    To examine how age sensitivity depends on the stellar population model and its ingredients, we compare the \hminus{} bump index among XSL, E-MILES, and C3K+MIST models generated using \texttt{FSPS} and M13 SSPs. In Figure~\ref{SSP}, we show these model families across age and metallicity at $\sigma = 300~\rm km/s$, including H$\delta$ and $\mathrm{D}4000$ as optical age features. In Figure~\ref{M13-SSP}, we show the \hminus{} bump and $\mathrm{D}4000$ indices at $R=100$ for XSL and M13. 
    
    We find broadly similar trends in H$\delta$ and $\mathrm{D}4000$ among the models, but substantial differences in the \hminus{} bump. Both XSL+PARSEC/COLIBRI and M13 predict a pronounced intermediate-age enhancement, while E-MILES shows only a weak age dependence.

    To isolate the effect of TP-AGB weighting, we compare two C3K+MIST models generated with \texttt{FSPS} v4.0\footnote{\url{https://github.com/cconroy20/fsps}}. Both use the empirical O-rich TP-AGB templates of \citet{Lancon-2002} (LW02; \texttt{use\_lw\_tpagb}=1), as in the reference TP-AGB-on mocks. Increasing the TP-AGB weight from \texttt{agb}=0 to 2, with other settings fixed, strengthens the intermediate-age \hminus{} bump. We use these contrasting weights to suppress or enhance the TP-AGB contribution relative to its fiducial value.
    
    The enhancement near a stellar age of $\sim1$~Gyr suggests that the \hminus{} bump could help trace recent quenching phases. By accounting for its dependence on the population model, we explore this possibility below.

\subsection{The  \texorpdfstring{$\mathrm{H}^{-}$}{H-} bump during the transition to quiescence}
    \label{ssec:sfh}

    We next ask how this intermediate-age sensitivity appears in populations undergoing quenching. For this test, we use XSL-based CSPs, which combine empirical O-rich and C-rich TP-AGB spectra with PARSEC/COLIBRI evolution \citep{Verro-2022a}.

\begin{figure}[htbp]            
    \includegraphics[width=\columnwidth]{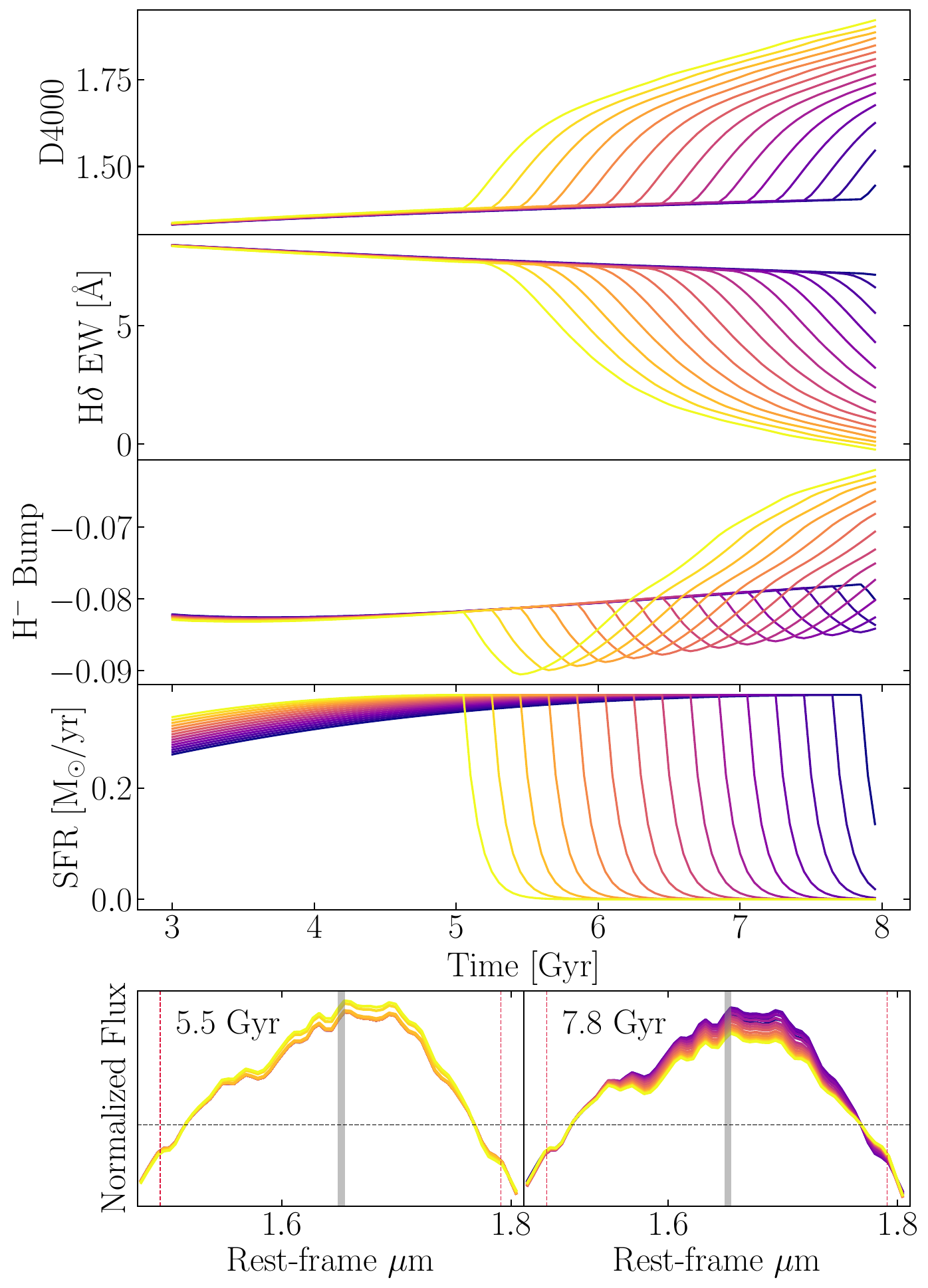}
    \caption{Top three panels: evolution in the XSL-based CSPs of $\mathrm{D}4000$, H$\rm \delta$, and the \hminus{} bump index with model time. All tracks use solar-metallicity ($[\mathrm{Fe/H}]=0$) XSL PARSEC/COLIBRI SSPs \citep{Verro-2022a}. The \hminus{} bump panel shows \(I_{\rm mag}\) in magnitudes; a stronger \hminus{} bump has a more negative value, so its transient enhancement appears as a downward trough. The fourth panel shows the corresponding SFHs with the same colors. The SFHs differ in their quenching onset times, $t_{\rm q}$. The two panels below show enlarged views of the normalized spectra around 1.6 \si{\micro\metre} at model times 5.5 Gyr and 7.8 Gyr, illustrating the \hminus{} bump at two observation times. The vertical red lines mark the sidebands used to construct the pseudo-continuum, and the gray line shows the wavelength center of this feature.}
\label{csp-tq-indices}
\end{figure}

    We first vary the quenching onset time at solar metallicity ($[\mathrm{Fe/H}]=0$), keeping $\tau_q=0.1$~Gyr fixed (Figure~\ref{csp-tq-indices}). After quenching begins, $\mathrm{D}4000$ rises and stellar H$\delta$ absorption declines. The \hminus{} bump instead strengthens and then weakens, reaching its greatest strength approximately 0.3~Gyr after the onset of quenching. Its strength therefore depends on when we observe the population during this transition.

\begin{figure}[htbp]            
    \includegraphics[width=\columnwidth]{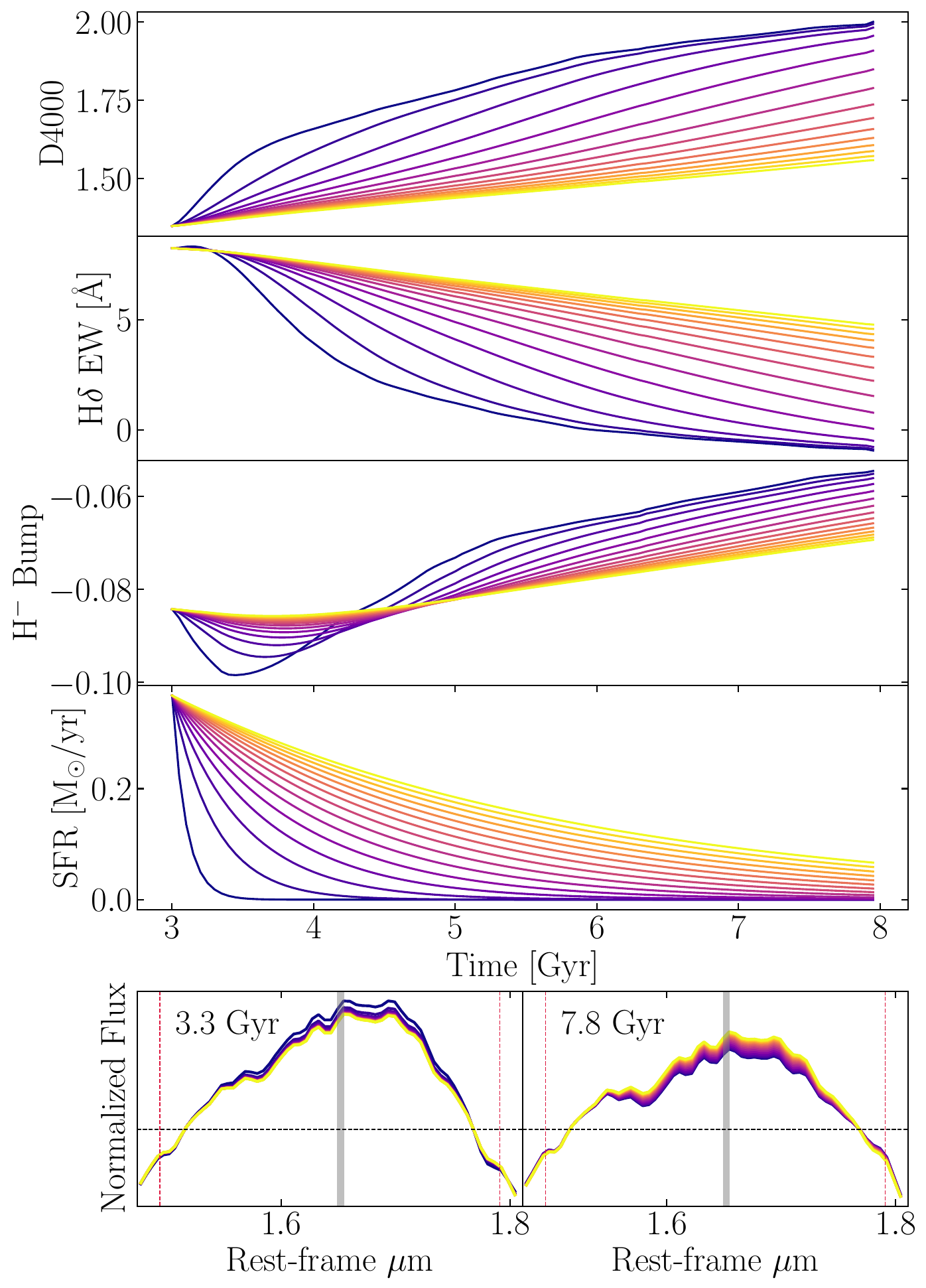}
    \caption{Same as Figure \ref{csp-tq-indices}, but with fixed $t_q = 3~\rm Gyr$ and varying $\tau_q$. The bottom panels show normalized spectra at model times 3.3 and 7.8 Gyr.}
\label{csp-tauq-indices}
\end{figure}

    To examine the effect of the quenching timescale, we then fix $t_q=3$~Gyr and vary $\tau_q$ from 0.1 to 3~Gyr (Figure~\ref{csp-tauq-indices}). Faster quenching produces a more pronounced transient \hminus{} bump enhancement, consistent with young stellar light fading rapidly and exposing the intermediate-age population. We show the corresponding changes in the normalized near-infrared spectra in the lower panels of both figures.

\subsection{Testing TP-AGB contributions and complementarity with optical indices}
    \label{sec:index}

    We now use the paired FSPS populations of Section~\ref{ssec:fsps_mocks} to test how TP-AGB stars affect the \hminus{} bump and whether it provides information beyond $\mathrm{D}4000$ and H$\delta_{\rm A}$. With FSPS, we can remove the TP-AGB contribution while keeping the SFHs and metallicities identical. This is a controlled experiment that the XSL SSP library with fixed TP-AGB contributions does not allow.

\begin{figure*}[t]
    \centering
    \includegraphics[width=\textwidth]{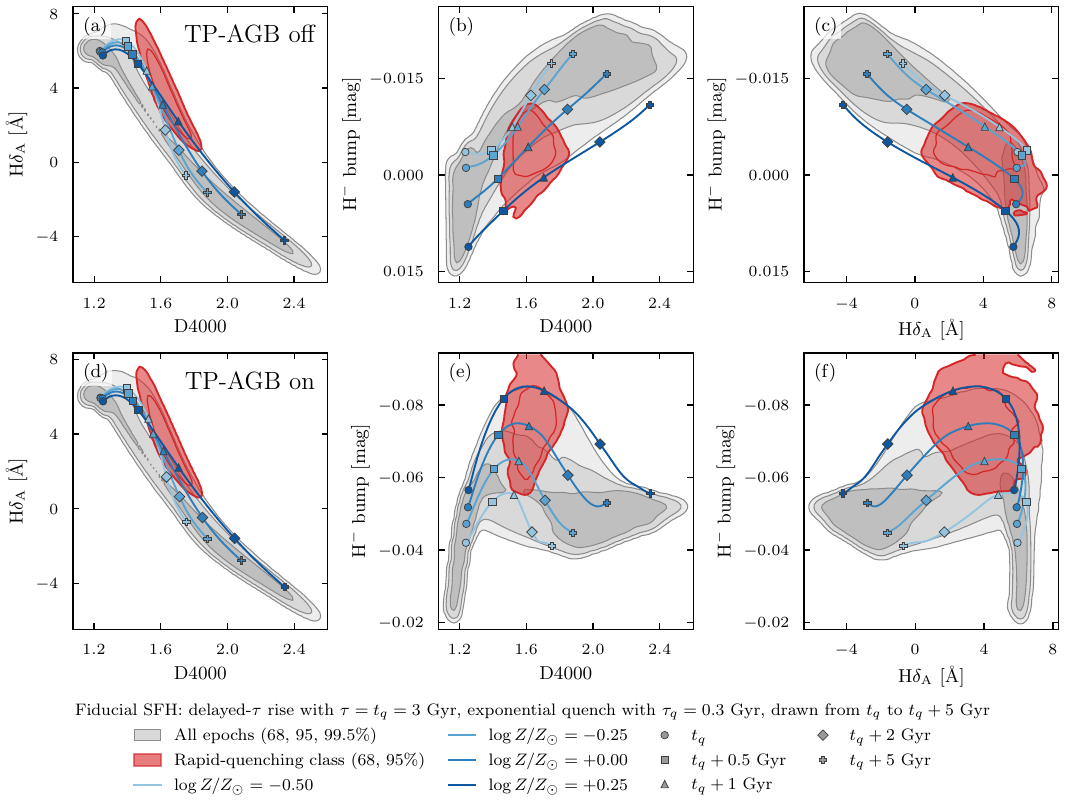}
    \caption{Spectral-index planes for the reference FSPS ensemble with TP-AGB light removed (top row) or enhanced using empirical O-rich templates (bottom row), as defined in Section~\ref{ssec:fsps_mocks}. Columns show H$\delta_{\rm A}$ versus $\mathrm{D}4000$, the \hminus{} bump versus $\mathrm{D}4000$, and the \hminus{} bump versus H$\delta_{\rm A}$. \hminus{} bump axes are inverted: more negative indices indicate a stronger \hminus{} bump. Gray contours enclose 68, 95, and 99.5\% of the 482\,000 retained sampled epochs, and red contours enclose 68 and 95\% of the rapid-quenching class. Blue curves follow histories with $t_q=3$~Gyr and $\tau_q=0.3$~Gyr at four metallicities, from $t_q$ to $t_q+5$~Gyr; symbols mark the elapsed times indicated in the legend. The optical indices use the $\sigma=300~\mathrm{km\,s^{-1}}$ spectra, and the \hminus{} bump uses the $R=100$ instrumental smoothing described in the text. The optical distributions change little between configurations, whereas the \hminus{} bump distinguishes their TP-AGB contributions and places the TP-AGB-on rapid-quenching class toward the strongest features. The two rows use different ranges on the \hminus{} bump axes.}
    \label{fig:index_planes_agb_on_off}
\end{figure*}

    We first compare the TP-AGB-off and TP-AGB-on populations described in Section.\ref{ssec:fsps_mocks} on three spectral-index planes (Figure~\ref{fig:index_planes_agb_on_off}). The optical plane changes little, whereas both planes containing the \hminus{} bump change substantially. With TP-AGB on, the rapid-quenching class lies toward the strongest \hminus{} bump values and has a median index of $-0.075$~mag, compared with $-0.004$~mag with TP-AGB off. This contrast suggests that we can use the region to test TP-AGB prescriptions. The rapid quenching phase, as illustrated in red, covers the strongest \hminus{} bump region, while overlapping with the remaining populations. We also find that the metallicity trend reverses between configurations.

   Using the same empirical O-rich LW02 spectra but varing the weight \texttt{agb} gives median \hminus{} bump indices of $-0.004$, $-0.036$, and $-0.059$~mag for \texttt{agb}=0, 1, and 2, when $1.3\leq\mathrm{D}4000<1.5$. The \hminus{} bump is therefore already substantial at \texttt{agb}=1 and strengthens as the TP-AGB contribution increases. For comparison, setting \texttt{use\_lw\_tpagb}=0 gives a much weaker median index of $-0.009$~mag. This configuration retains TP-AGB light but assigns these stars the same C3K atmosphere spectra as ordinary giants with the same temperature, gravity, and composition. It lacks a distinct spectral treatment of TP-AGB atmospheres, including the stronger molecular absorption that lowers the pseudo-continuum in the empirical templates. Its weak \hminus{} bump therefore illustrates a limitation of the spectral treatment.

\begin{figure*}[t]
    \centering
    \includegraphics[width=\textwidth]{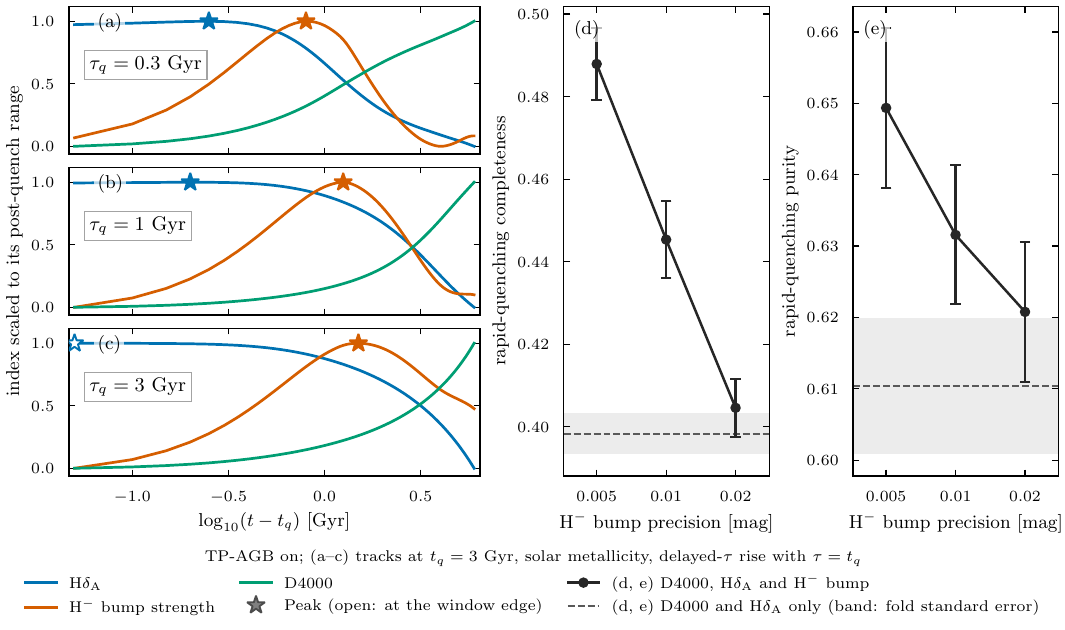}
    \caption{Different temporal responses of the indices and the information added by the \hminus{} bump in the TP-AGB-on model. Panels (a)--(c) show solar-metallicity histories with $t_q=\tau=3$~Gyr and $\tau_q=0.3$, 1, and 3~Gyr. H$\delta_{\rm A}$ (blue), \hminus{} bump strength $-I_{\rm mag}$ (vermilion), and $\mathrm{D}4000$ (green) are each scaled to their own range over 0.05--6~Gyr after quenching begins, illustrating their timing. Stars mark maxima of H$\delta_{\rm A}$ and \hminus{} bump strength; the open star marks a maximum at the first sampled epoch. Panels (d) and (e) show completeness and purity for the rapid-quenching class using a 25-nearest-neighbor classifier. Points use all three indices; dashed lines use only the two optical indices. Gaussian noise has standard deviations of 0.05 in $\mathrm{D}4000$, 0.5~\AA\ in H$\delta_{\rm A}$, and the indicated precision in the \hminus{} bump. Distances are scaled by these uncertainties. Validation uses five folds grouped by SFH and three noise realizations. Error bars and shaded bands show the fold scatter divided by $\sqrt{5}$, averaged over those realizations.}
    \label{fig:age_sensitivity_classifier}
\end{figure*}

    We then compare the temporal responses of the three indices to examine how the \hminus{} bump might complement the optical diagnostics (Figure~\ref{fig:age_sensitivity_classifier}(a)--(c)). In these FSPS histories, the \hminus{} bump reaches its greatest strength after H$\delta_{\rm A}$ reaches its maximum. Its transient enhancement also weakens as quenching becomes slower, in qualitative agreement with the XSL results. The detailed timing and amplitudes remain model-dependent. We normalize each curve to compare response timing and remove amplitude differences from the plot. In Figure~\ref{fig:age_sensitivity_classifier}, the actual \hminus{} bump excursion declines from 0.023 to 0.006~mag between $\tau_q=0.3$ and 3~Gyr.

    To test whether these different responses help us identify rapid quenching, we compare classification using only $(\mathrm{D}4000,\mathrm{H}\delta_{\rm A})$ with classification that also includes the \hminus{} bump. We use a 25-nearest-neighbor classifier with unweighted majority voting across the four classes. We add independent Gaussian measurement noise, scale index differences by the assumed errors, and use the same noisy optical measurements in both tests. We divide the SFHs into five validation folds, keeping all epochs of a given history together so that it cannot enter both training and validation. We retain the sampled class frequencies and repeat the test for three noise realizations. We assess the results using completeness, the fraction of true rapid-quenching epochs recovered, and purity, the fraction of selected epochs that belong to that class.

    Adding the \hminus{} bump improves both completeness and purity, with larger gains at higher measurement precision (Figure~\ref{fig:age_sensitivity_classifier}(d)--(e)). At 0.005~mag precision, completeness increases from 0.398 to 0.488 and purity from 0.610 to 0.649. At 0.010~mag, they reach 0.445 and 0.632, while at 0.020~mag the gain is small. Repeating the test with the TP-AGB-off reference mocks yields no comparable improvement in completeness.

    We also test whether the \hminus{} bump improves estimates of the elapsed time since quenching and the quenching timescale. For this purpose, we use a 25-nearest-neighbor regression that predicts each logarithmic target by averaging over its nearest training epochs. We apply the same noise and validation scheme to epochs within 6~Gyr after quenching onset. At 0.005~mag precision, adding the \hminus{} bump reduces the RMS errors on log elapsed time and log quenching timescale from 0.240 to 0.221 dex and from 0.311 to 0.298 dex, respectively. These improvements show that the \hminus{} bump adds information within the FSPS mocks.

    To test whether these results depend on our choice of SFH, we present three alternative FSPS history families in Appendix~\ref{sec:sfh_families}. We find the same association between a strong \hminus{} bump and rapid quenching in each family.

\section{Discussion} 
    \label{sec:discuss}

\subsection{Limitations of SSP Models}
\label{sec:limit}

    The TP-AGB contribution depends on the evolutionary lifetimes and weights of these stars, as well as their assigned spectra. The comparison between XSL and E-MILES in Figure~\ref{SSP} involves differences in both ingredients. Under the same isochrone, the FSPS experiments show the variance of including TP-AGB spectra.

    The X-shooter Spectral Library and PARSEC/COLIBRI isochrones represent a relatively TP-AGB-strong modeling framework. COLIBRI includes dredge-up, hot-bottom burning, pulsation, and mass loss, calibrated using Magellanic Cloud cluster star counts and luminosity functions \citep{bressan2012parsec, marigo2013evolution, marigo2017new}. The Padova00 isochrones \citep{girardi2000evolutionary} adopt more simplified synthetic prescriptions for TP-AGB evolution, leading to different predictions for TP-AGB lifetimes and their contributions to the integrated NIR light. MIST follows stellar evolution with MESA \citep{dotter2016mesa}, but its TP-AGB properties are not explicitly calibrated against observed populations. M13 adopts a fuel-consumption approach, with TP-AGB contributions recalibrated using Magellanic Cloud star clusters \citep{Maraston-2005a, Maraston-2005b, Noel-2013}. We find that the non-monotonic age response of the \hminus{} bump relies on TP-AGB contributions to SSP models.

    The metallicity trend also reflects the template assignment, and metallicity variation in TP-AGB models is not well constrained. Therefore, we should be careful when interpreting the metallicity evolution across the intermediate ages. In the fiducial FSPS tracks, the \hminus{} bump weakens with increasing metallicity when TP-AGB light is removed, but strengthens when the empirical TP-AGB component is included. FSPS uses metallicity-dependent temperature labels to select among fixed spectral shapes, and assigns them only to O-rich TP-AGB stars with $\log_{10}(T_{\rm eff}/\mathrm{K})<3.6$. The trend therefore depends on the MIST temperature distribution, this assignment rule, and the spectra. Moreover, the adopted stellar population grids contain no carbon-rich TP-AGB stars over the metallicity range used for the mocks. Testing C-rich populations and alternative temperature scales remains necessary before interpreting the \hminus{} bump as a quantitative measure of the total TP-AGB contribution.

    Empirically calibrated TP-AGB prescriptions remain subject to systematic uncertainties because their predicted evolutionary lifetimes and number counts depend on the adopted physical prescriptions and calibration samples \citep[e.g.,][]{Maraston-2005a, Maraston-2005b, Noel-2013}. For example, \cite{girardi2013insidious} showed that Magellanic Cloud clusters at ages around 1.6 Gyr coincide with a ``TP-AGB boosting'' period, when stars with slightly different initial masses simultaneously enter the TP-AGB phase due to rapid changes in core He-burning lifetimes. This temporary enhancement can bias empirical calibrations if interpreted as representative of all intermediate-age populations. As implied by several studies \citep[e.g.][]{Kriek-2010, Zibetti-2013}, the contribution of TP-AGB stars to the NIR light of galaxies may be weaker than predicted by some models. Measurements of the \hminus{} bump in populations with independently constrained ages and metallicities can help distinguish these stellar population prescriptions. Comparisons with large observational samples can therefore serve two purposes: testing whether the predicted temporal behavior of the \hminus{} is realized in galaxies, and constraining the treatment of TP-AGB stars in stellar population models. In this sense, systematic discrepancies between the predicted and observed \hminus{} can provide population-level indications of uncertainties in TP-AGB modeling.

\subsection{Comparison with Observations}
    \label{sec:comparison}

    JWST/NIRSpec PRISM spectroscopy already allows the \hminus{} bump and $\mathrm{D}4000$ to be measured together at intermediate redshift. Its resolution corresponds to roughly 40~\AA{} near rest-frame 4000~\AA{}, comparable to the H$\delta$ central bandpass and therefore insufficient for accurate H$\delta$ measurements. $\mathrm{D}4000$ is more robust than H$\delta$ because it is a broad continuum-break index, yet the low resolution means that directly integrating over several pixels to derive the continuum flux leads to large fluctuations. Here we measure $\mathrm{D}4000$ by linearly fitting the blue and red continuum windows in frequency space and using the ratio of analytic integrals from these two linear functions as the index. We estimate $\mathrm{D}4000$ and the \hminus{} bump uncertainties at the 16th and 84th percentiles from 1000 Monte Carlo realizations. 
    
    We select publicly available JWST NIRSpec MSA PRISM spectra based on the Dawn JWST Archive (DJA) Public NIRSpec datasets v4.4 \citep{deGraaff-2025, Heintz-2024}. We choose quiescent galaxies instead of star-forming ones to avoid potential contamination of stellar indices by non-stellar components (e.g., nebular emission lines and AGN features, which have not been considered in our models). We adopt the UVJ selection criteria for $1<z<4$ galaxies in \cite{Muzzin-2013}. We directly use the rest-frame UVJ colors from the best-fitting EAZY photometric template combination \citep{Valentino-2023} in the catalog, requiring robust redshift measurements $\rm{grade} == 3$ and $1.1 < z_{\rm{best}}<1.8$ to cover $\mathrm{D}4000$ and the \hminus{} bump features simultaneously. We also require the median signal-to-noise ratio to be higher than 50, resulting in a preliminary sample of 56 quiescent galaxies. We then exclude cases with problematic spectral quality, crowded fields, and bright neighbor contamination after visual inspection, and ensure that none of the remaining spectra have emission lines. The final sample has 47 targets in total, from JWST Program 1180, 1181, 1286\citep{Eisenstein-2026, Curtis-Lake-2026, Scholtz-2026}, 1208\citep{Willott-2022}, 1211, 1214\citep{Maseda-2024},  1345\citep{Finkelstein-2023}, 2198\citep{Barrufet-2025}, 2565\citep{Nanayakkara-2025}, 4233\citep{deGraaff-2025}, 5224\citep{Naidu-2026} and 6368\citep{Kokorev-2025}. 

    Because most of the 47 galaxies are extended, extracting 1D spectra should be handled carefully to recover most of the stellar continuum and avoid self-subtraction issues. We use the public DJA global sky background\footnote{\url{https://github.com/gbrammer/msaexp/tree/main/msaexp/data/msa_sky}} if available \citep{deGraaff-2025}.  We processed the raw exposures through Stage 1 of the STScI \texttt{jwst} pipeline \citep[v1.17.1,][]{Bushouse2025zndo..14597407B}, adopting CRDS context \texttt{jwst\_1322.pmap}. Subsequent calibration was performed with \texttt{msaexp} \citep[0.1.dev1+g44392ac,][]{Brammer2023zndo...8319596B}, including 1/f-noise correction and read-noise variance rescaling. The spectra were then background-subtracted using visit-matched global-sky templates, combined across individual exposures, and optimally extracted.

    There are 22 targets that are not released with a master-sky spectrum. To avoid self-subtraction of these extended sources, we use the publicly available, boxcar-extracted 1D spectra from the JADES GTO pipeline \citep{Scholtz-2026}\footnote{\url{https://jades.herts.ac.uk/jwst_archive}} instead. They are reduced with the ``nod2'' background subtraction, which employs only the two outer nod positions. This method combines the accuracy of nod subtraction without introducing any self-subtraction. This procedure discards the central nod position and obtains the background from a single nod position, resulting in shorter exposure time and higher background noise. Here we focus only on high signal-to-noise targets, so this shortcoming has negligible effect on our index measurements.

\begin{figure*}[t]
    \centering
    \includegraphics[width=\textwidth]{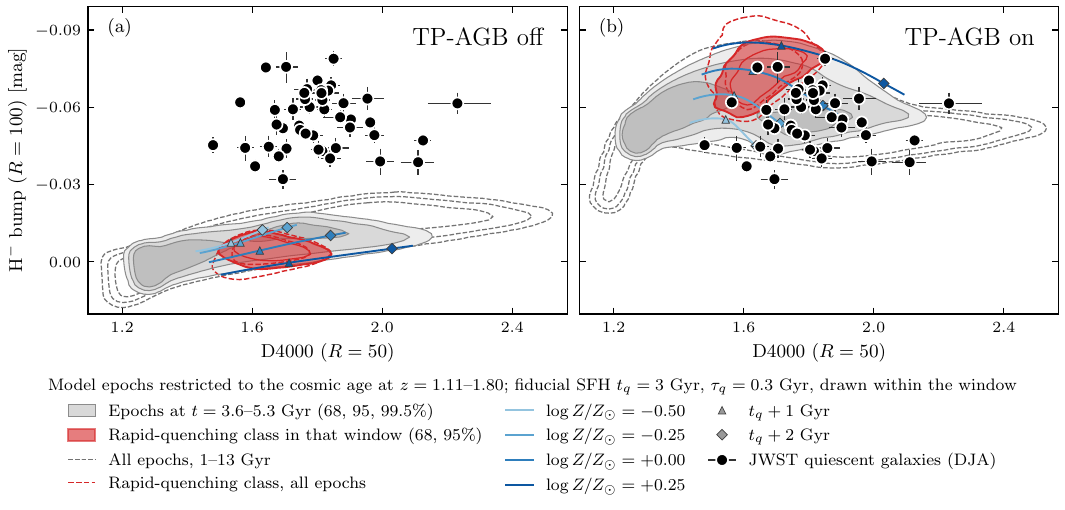}
    \caption{JWST measurements compared with the reference FSPS mocks in the $\mathrm{D}4000$ versus \hminus{} bump plane. We change the evolution time from 13 Gyr to 6 Gyr to match the cosmic ages of the samples. The panels show TP-AGB off (left) and TP-AGB on (right) on a common \hminus{} bump scale. Black points represent the 47 quiescent galaxies at $1.1<z<1.8$ selected from public NIRSpec/MSA PRISM spectra. Horizontal error bars span the 16th--84th percentile bounds of $\mathrm{D}4000$. Contour fractions and fiducial metallicity tracks follow Figure~\ref{fig:index_planes_agb_on_off}, but here model $\mathrm{D}4000$ is remeasured with $R=50$, and the \hminus{} bump retains the $R=100$ smoothing to approximate the NIRSpec PRISM observed-frame resolution. Four of those galaxies are within the 95 percent contour of the rapid-quenching class, and two are within its 68 percent contour. The observation distributions broadly overlap the TP-AGB-on distribution more than the TP-AGB-off one.}
    \label{2Dobs}
\end{figure*}

    
    Figure~\ref{2Dobs} compares the observations with the paired FSPS populations. The measured $\mathrm{D}4000$ values span 1.48 -- 2.23 and the \hminus{} bump indices $-0.079$ to $-0.032$~mag. Within this $\mathrm{D}4000$ range, the sampled TP-AGB-off ensemble reaches only $-0.023$~mag. Every observed central value indicates a stronger \hminus{} bump. The TP-AGB-on configuration broadly overlaps the observations. The separation provides strong preliminary observational evidence supporting the TP-AGB-enhanced scenario for these galaxies, though further validation with different stellar libraries, TP-AGB weight, and evolutionary pathways is needed to draw a solid conclusion. Among these samples, 34 of the 47 galaxies lie within its 99.5 percent contour, four within the 95 percent contour of the rapid-quenching class, and two within its 68 percent contour. However, we note that the index comparison is illustrative, given that this overlap depends strongly on the adopted population model and SFHs.  As we further discuss in Appendix~\ref{msa-reduction}, the observational measurements of indices with the global-sky and ``nod2'' approaches also show discrepancies due to different data reduction procedures. Nevertheless, the values from these two methods under the same sample are both consistent with the ``TP-AGB-on'' model.

\begin{figure*}[t]
    \centering
    \includegraphics[width=\textwidth]{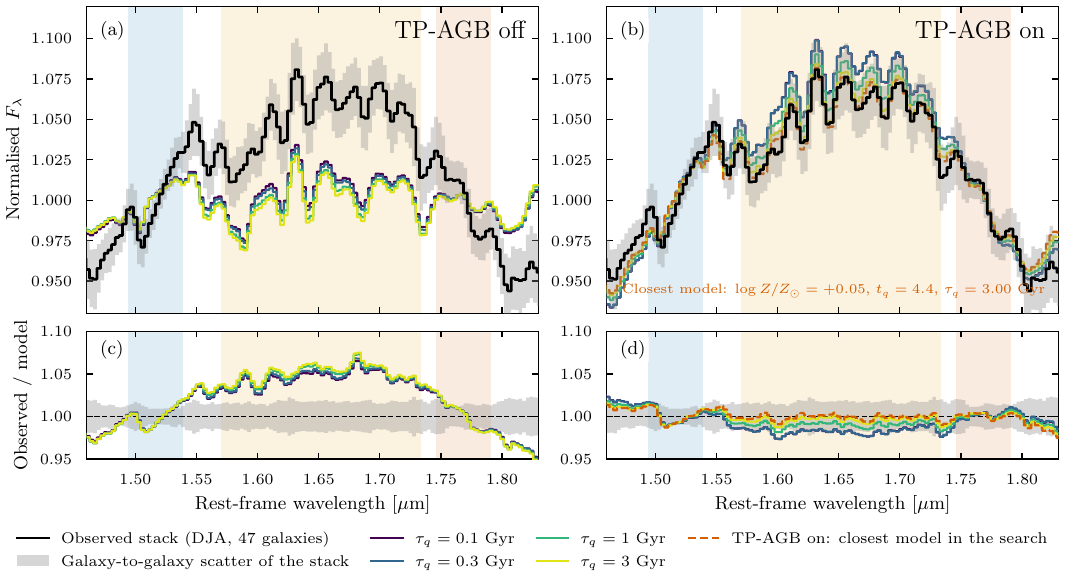}
    \caption{The stacked JWST \hminus{} bump compared with FSPS spectra with TP-AGB light removed (left) or enhanced empirical TP-AGB templates (right), as defined in Section~\ref{ssec:fsps_mocks}. Each observed and model spectrum is divided by a straight line fitted to all pixels in the two Hbump\_opt pseudo-continuum windows. The black curve is the signal-to-ratio-weighted mean of the 47 normalized spectra. Gray shading shows the weighted galaxy-to-galaxy scatter. The models are evaluated at a single epoch, the cosmic age at the median redshift, and forward-modeled to the data. For each galaxy, the model is smoothed to the NIRSpec PRISM spectral resolution at the observed wavelength of its bump and combined with the S/N weights of the observed stack. Colored curves have solar metallicity, $t-t_q=1$~Gyr, and $\tau_q=0.1$, 0.3, 1, and 3~Gyr. The dashed histogram in panel (b) is the closest TP-AGB-on model to the observed stack in a grid search at the same redshift, ranked by the RMS of the residual in units of the galaxy-to-galaxy scatter over 1.494--1.791~$\mu$m. Its labeled parameters describe one grid model. The lower panels show observed/model ratios, and gray shading shows the fractional scatter. Blue, yellow, and red shading mark the pseudo-continuum and feature bands.}
    \label{obs-stack}
\end{figure*}

    Figure~\ref{obs-stack} compares the normalized observed stack with FSPS spectra at an epoch representative of the sample. At fixed time since quenching, the illustrated TP-AGB-off models remain too weak across the feature band as $\tau_q$ varies from 0.1 to 3~Gyr. The TP-AGB-on models reproduce the broad profile, with the 1 and 3~Gyr examples lying within the galaxy-to-galaxy scatter across the feature band. We search over metallicity and quenching parameters to identify the closest TP-AGB-on spectrum, shown as a dashed line. We also tested on XSL-based CSPs selected at similar index values. Both of them give comparably good agreement with the stacked spectra.

    Existing models with substantial TP-AGB contribution can thus describe the overall shape of the broad \hminus{} bump, although reproducing other TP-AGB-sensitive features remains uncertain. For example, \cite{Lu-2026} demonstrate that M13 \citep{Noel-2013} generally provides better overall fits to 27 quiescent galaxy samples, including \hminus{} bump features, while it remains challenging for all models to reproduce the CN absorption feature at 0.92 $\mu$m. Observational measurements of the \hminus{} bump and other TP-AGB-sensitive features in larger, more complete samples can provide further constraints on stellar population models. 

\subsection{Observational Prospects}
    \label{ssec:forecast}
    
    Unlike narrow absorption features such as H$\delta$, the \hminus{} continuum feature that is centered at 1.6 $\mu m$ extends over a broad wavelength range ($\sim0.3~\mu$m). As shown above, the feature remains detectable after degrading the spectra to $R\sim50$, making it significantly less sensitive to spectral resolution than traditional age indicators. This property enables measurements from low-resolution spectroscopy and, in some cases, even from photometric observations.

    NIRSpec PRISM covers approximately 0.6--5.3~$\mu$m at $R\sim30$--$300$, allowing joint measurements of $\mathrm{D}4000$ and the \hminus{} bump at intermediate redshifts up to approximately $z=2$, with both pseudo-continuum bands remaining covered. Spatially resolved spectroscopy could also probe radial variations in the \hminus{} bump in sufficiently bright, resolved galaxies.

    The all-sky SPHEREx survey \citep{Dore-2016, Dore-2018} offers complementary data at low redshift. Studies using simulated SPHEREx data identify the 1.6~\si{\micro\metre} region as useful for stellar mass estimation \citep{Lee-2025}, while early observations reveal discrepancies with SSP predictions \citep{Lee-2026}. The survey's large samples could reveal statistical trends with stellar mass, environment, and star-formation state, once measurement biases are characterized.

    Euclid \citep{EuclidCollaboration-2025} and Roman \citep{Akeson-2019, Spergel-2015} slitless spectra could also probe the \hminus{} bump at low redshift, where their wavelength coverage includes the full index. Broad-band colors from these surveys or JWST/NIRCam may provide indirect indicators where filters bracket the redshifted \hminus{} bump, but require spectroscopic calibration and depend more strongly on redshift estimation, dust attenuation, and population modeling.

\subsection{Caveats}
    \label{ssec:caveats}

    Our alternative SFHs give qualitatively similar results, but they cover only a limited range of possible histories. Real galaxies can undergo bursts, rejuvenation, and repeated suppression, particularly at high redshift and low mass \citep[e.g.,][]{Asada-2024, Perry-2025, Mintz-2026}, where star formation can be stochastic \citep{Caplar-2019, Tacchella-2020, Carvajal-Bohorquez-2025}. Tests with non-parametric SFHs \citep[e.g.,][]{Leja-2019} would help establish whether the \hminus{} bump adds information for less restrictive histories. Our classifier and regression use the same population model in training and validation and omit correlated observational errors and model mismatch. The improvements in these tests therefore do not establish the uncertainties we can achieve for observed quenching histories.
    
    The mocks retain circumstellar AGB dust but omit interstellar attenuation, nebular emission, and AGN light. Hot-dust emission can dilute the \hminus{} bump, especially in obscured starbursts and AGN hosts. Dust attenuation on the other side should affect the \hminus{} bump less strongly than optical age indicators. 

    High-order Brackett lines and [Fe II] at 1.644~\si{\micro\metre} lie in the \hminus{} bump region \citep{Morel-2002, deGrijs-2004}. The broad feature band reduces the contribution of individual weak lines compared with narrow absorption indices. Emission from young stars declines after quenching, but AGN and shocks can still contribute. Lines in the feature band strengthen the apparent \hminus{} bump, while contamination in the pseudo-continuum bands changes the reference level. We omit these effects here, but they will need to be assessed for galaxies with significant emission.

    Flux calibration, spectral sampling, and pseudo-continuum placement can also shift an index whose useful variations are only a few hundredths of a magnitude. A single index discards details of the spectral shape and neighboring features. Full optical and near-infrared spectral fitting may use this information to improve quenching constraints. However, flexible continuum corrections, such as high-order polynomials, can remove broad spectral structure and reduce NIR age sensitivity \citep{Baldwin-2018}. Their effect on the \hminus{} bump should therefore be assessed alongside flux-calibration and stellar-population uncertainties.

\section{Conclusion}
    \label{sec:conclude}

    In this first paper, we investigate the 1.6~\si{\micro\metre} \hminus{} bump as a diagnostic of intermediate-age stellar populations and recent galaxy quenching. Our main findings are:

    \begin{enumerate}
    
    \item The \hminus{} bump's intermediate-age enhancement varies strongly among stellar population models, and it depends strongly on the treatment of TP-AGB stars. Within the XSL and FSPS frameworks, with TP-AGB spectra incorporated, the \hminus{} bump index shows an enhancement in intermediate-age stellar populations. We propose an optimized index definition of the \hminus{} bump based on XSL SSP models, Hbump\_opt, to highlight its age-sensitive information.

    \item The FSPS mocks show that we can use the \hminus{} bump to test both the contribution of TP-AGB stars and the spectra assigned to them. With empirical O-rich templates at twice the fiducial TP-AGB weight, the rapid-quenching class has a median index of $-0.075$~mag, compared with $-0.004$~mag when TP-AGB light is removed. The optical index distributions change little. 

    \item XSL with PARSEC/COLIBRI predicts a transient \hminus{} bump enhancement that is more pronounced for faster quenching. The FSPS models with empirical templates show qualitatively similar behavior, although the timing and amplitude remain model-dependent. In the FSPS tests, including it alongside $\mathrm{D}4000$ and H$\delta_{\rm A}$ increases rapid-quenching completeness from 0.40 to 0.49 and purity from 0.61 to 0.65, at 0.005~mag \hminus{} bump precision. Parameter-recovery tests also yield modest reductions in the errors on elapsed quenching time and timescale. Alternative SFHs preserve the association between rapid quenching and a strong \hminus{} bump. These gains show that the additional information is present within the adopted models. How much it improves constraints on observed galaxies remains to be established.

    \item On the $\mathrm{D}4000$ -- \hminus{} bump plane, current JWST spectra already offer an observational test. The 47 quiescent galaxies at $z=1$--2 have a stronger \hminus{} bump than the TP-AGB-free FSPS control can reproduce over the sampled parameter space. Their measured indices broadly overlap the predictions of the empirical-template configuration. This provides strong preliminary evidence under the TP-AGB-enhanced assumptions and demonstrates the diagnostic potential of the \hminus{} bump. Differentiating particular TP-AGB implementations is beyond the scope of this work.
    \end{enumerate}

    We emphasize that current optical and near-infrared spectra allow the \hminus{} bump to be used to test TP-AGB prescriptions and explore the recent quenching phases. Further investigation of different models and additional components (dust, nebulae, and metallicity) should be considered carefully in a broader context. Moreover, full spectral fitting may recover more information from this region than a small set of indices. In the future, larger samples and a broader range of stellar population models and SFHs are needed to determine how much these analyses improve quenching constraints when template uncertainties and measurement errors are included. 
    
    In the interest of transparency and reproducibility, we provide the data and the scripts for the paper figures in a public repository\footnote{\url{https://github.com/SunnySco/hbump.git}}.

\begin{acknowledgments}

    We acknowledge Stefano Carniani and Sandro Tacchella for their support. 
    We thank Shi-Ying Lu, Emanuele Daddi, and Claudia Maraston for sharing the insights from their work with us.

    (Some of) The data products presented herein were retrieved from the Dawn JWST Archive (DJA). DJA is an initiative of the Cosmic Dawn Center (DAWN), which is funded by the Danish National Research Foundation under grant DNRF140.
    
    PJL and SH acknowledge the support from the Ministry of Science and Technology of the People’s Republic of China (MOST) Grant No. 2023YFA1605600, the National Natural Science Foundation of China (NSFC) Grant No. 12273015, and NSFC Grant No. 12433003.

    FDE acknowledges support by the Science and Technology Facilities Council (STFC), by the ERC through Advanced Grant 695671 ``QUENCH'', and by the UKRI Frontier Research grant RISEandFALL.

\end{acknowledgments}

\vspace{5mm}
\appendix
\section{Construction of Composite Stellar Populations}
\label{sec:appendixA}

    For an SSP spectrum normalized per unit formed stellar mass, the CSP spectrum is
    \[
        F_{\lambda}^{\mathrm{CSP}}(t)=\int_0^t\psi(t')F_{\lambda}^{\mathrm{SSP}}(t-t',Z)\,dt',
    \]
    where $\psi(t')$ is the SFR and $t-t'$ is stellar age. For the XSL library, we evaluate this integral with custom code that combines SSP spectra according to the adopted SFH. We verified the implementation against FSPS by using the same SSP spectra and SFH in both calculations and obtained consistent CSP spectra.

    For the FSPS populations, we generate CSP spectra directly with the internal SFH routines of \texttt{FSPS v4.0}, accessed through \texttt{python-fsps}, using the population settings in Section~\ref{ssec:fsps_mocks}. The 50~Myr spacing specifies the epochs at which we evaluate the spectra. The paired TP-AGB configurations use a common surviving stellar mass, including living stars and remnants when defining sSFR.

    We broaden the FSPS spectra on a uniform log-wavelength grid with a 30~$\mathrm{km\,s^{-1}}$ step, convolving the flux per unit log wavelength. The C3K native resolving power is $R=3000$ over 0.3--1~\si{\micro\metre} and $R=500$ over 1--2.5~\si{\micro\metre}. We subtract the corresponding native Gaussian variance separately in the optical and near-infrared segments. The optical product has target $\sigma=300~\mathrm{km\,s^{-1}}$; for the \hminus{} bump the target variance is $300^2+[c/(2.355\times100)]^2$ in $(\mathrm{km\,s^{-1}})^2$. This defines the downgrade convention for the mixed-template spectra. For the JWST comparison in Figure~\ref{2Dobs}, we use $R=50$ for the $\mathrm{D}4000$ region and $R=100$ near the \hminus{} bump.

\section{The framework for selecting the \texorpdfstring{$\mathrm{H}^{-}$}{H-}\ bump index windows}
\label{sec:appendixB}

    We demonstrate the index-window selection framework with a two-stage random search. Each candidate is a six-parameter ordered window vector (B1, B2, F1, F2, R1, R2) that defines the blue pseudo-continuum, feature, and red pseudo-continuum bands. The windows must be ordered, lie within the specified hard wavelength limits, and meet minimum widths for the blue, feature, and red bands. To generate initial candidates for our mixed-resolution SSP search, we first assign the minimum band widths, then distribute the remaining wavelength interval using a Dirichlet random draw. This procedure efficiently samples valid ordered windows across the allowed interval. We enforce equal blue and red pseudo-continuum band widths to prevent unrealistic width differences.

\begin{figure*}[htbp]            
    \includegraphics[width=1\textwidth]{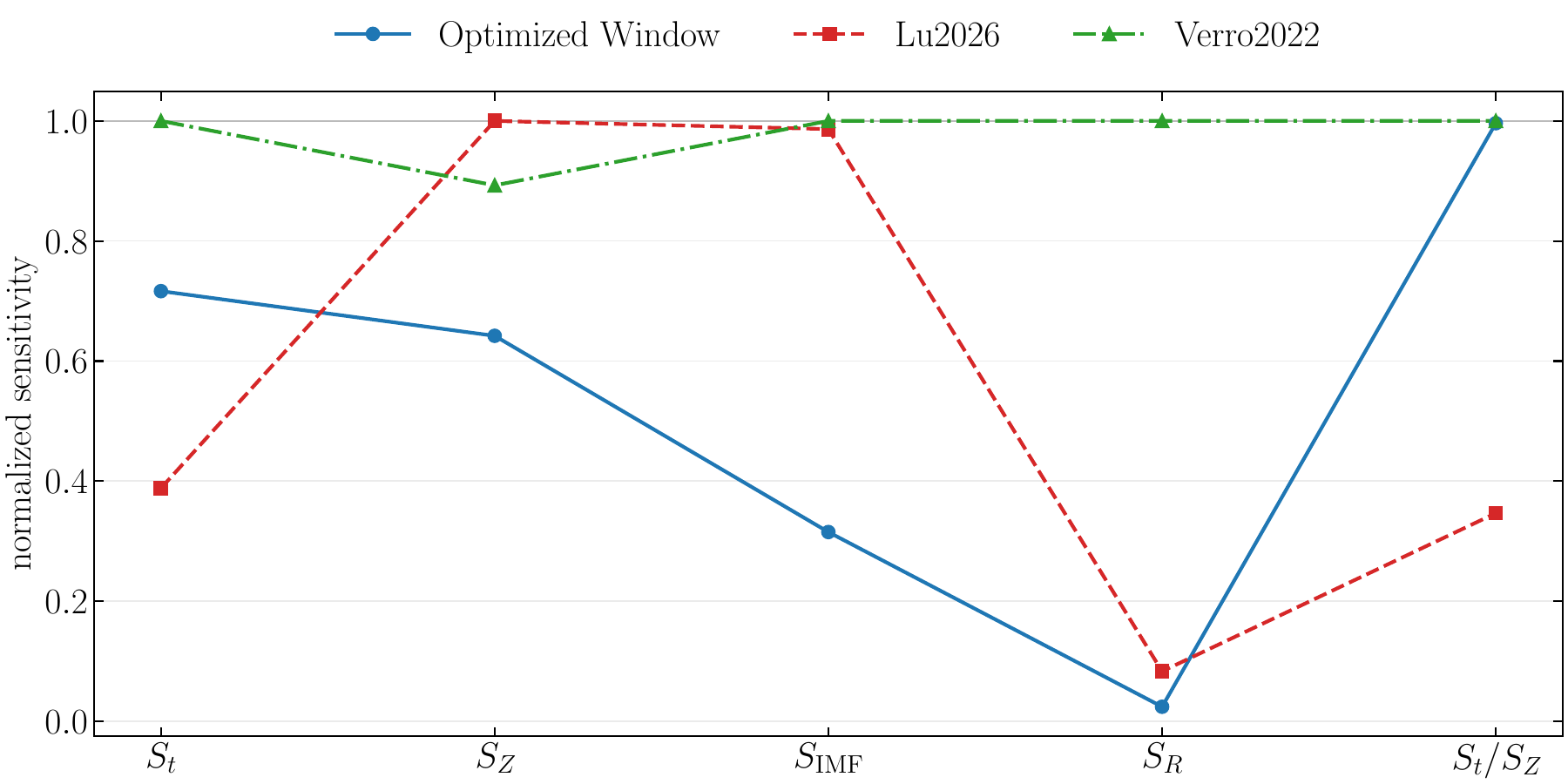}
    \caption{Comparison of the \hminus{} bump index sensitivity among the adopted Hbump\_opt definition, Verro2022 \citep{Verro-2022a}, and Lu2026 \citep{Lu-2026}. ``Optimized Window'' labels the adopted Hbump\_opt definition. For each sensitivity term on the x-axis, the scores are normalized by the maximum value among the three definitions. Hbump\_opt has lower age sensitivity than Verro2022 but a similar age-to-metallicity sensitivity ratio and reduced sensitivity to the other tested factors within the adopted model grid.}
\label{sens_plot}
\end{figure*}

    We rank the initial candidates by the objective function $\mathcal{L}$ described in Section~\ref{sec:hbump}. In the second stage, we draw Gaussian perturbations around the top-ranked candidates and reevaluate those that remain valid and finite. The highest-scoring candidate is the Hbump\_opt definition adopted in Section~\ref{sec:hbump}. Figure \ref{sens_plot} illustrates the resulting tradeoff: the adopted definition has greater age sensitivity than Lu2026, lower age sensitivity than Verro2022, a similar age-to-metallicity sensitivity ratio to Verro2022, and reduced sensitivity to the other tested factors under the definitions. This framework can be adapted by changing the objective, parameter ranges, and window constraints to suit other models, scientific goals, and observing conditions.

\section{Dependence on the Star-Formation History}
\label{sec:sfh_families}

    We test whether the association between the \hminus{} bump and rapid quenching depends on the reference SFH by considering three alternatives with the same TP-AGB-on configuration, metallicity draws, observation epochs, and classification rules as in Section~\ref{ssec:fsps_mocks}. In the first, the post-quenching SFR declines linearly to zero,
    \[
    \psi(t\geq t_q)=e^{-1}\max\!\left[0,1-\frac{t-t_q}{\Delta_q}\right],
    \qquad \Delta_q=2\ln2\,\tau_q,
    \]
    so its half-life equals that of the reference exponential decline. In the second, star formation is truncated abruptly at $t_q$. Both retain $\tau=t_q$ before quenching. In the third, the delayed-$\tau$ timescale is drawn independently from a log-uniform distribution over 0.5--5~Gyr, and an exponential decline is joined continuously at $t_q$. This third family therefore changes the pre-quenching history and its connection to the subsequent decline, and $t_q$ need not coincide with the SFR maximum. The same draws of $t_q$, $\tau_q$, and metallicity are used where applicable, but $\tau_q$ has no role in the abrupt-truncation model.

\begin{figure*}[p]
    \centering
    \includegraphics[width=0.94\textwidth]{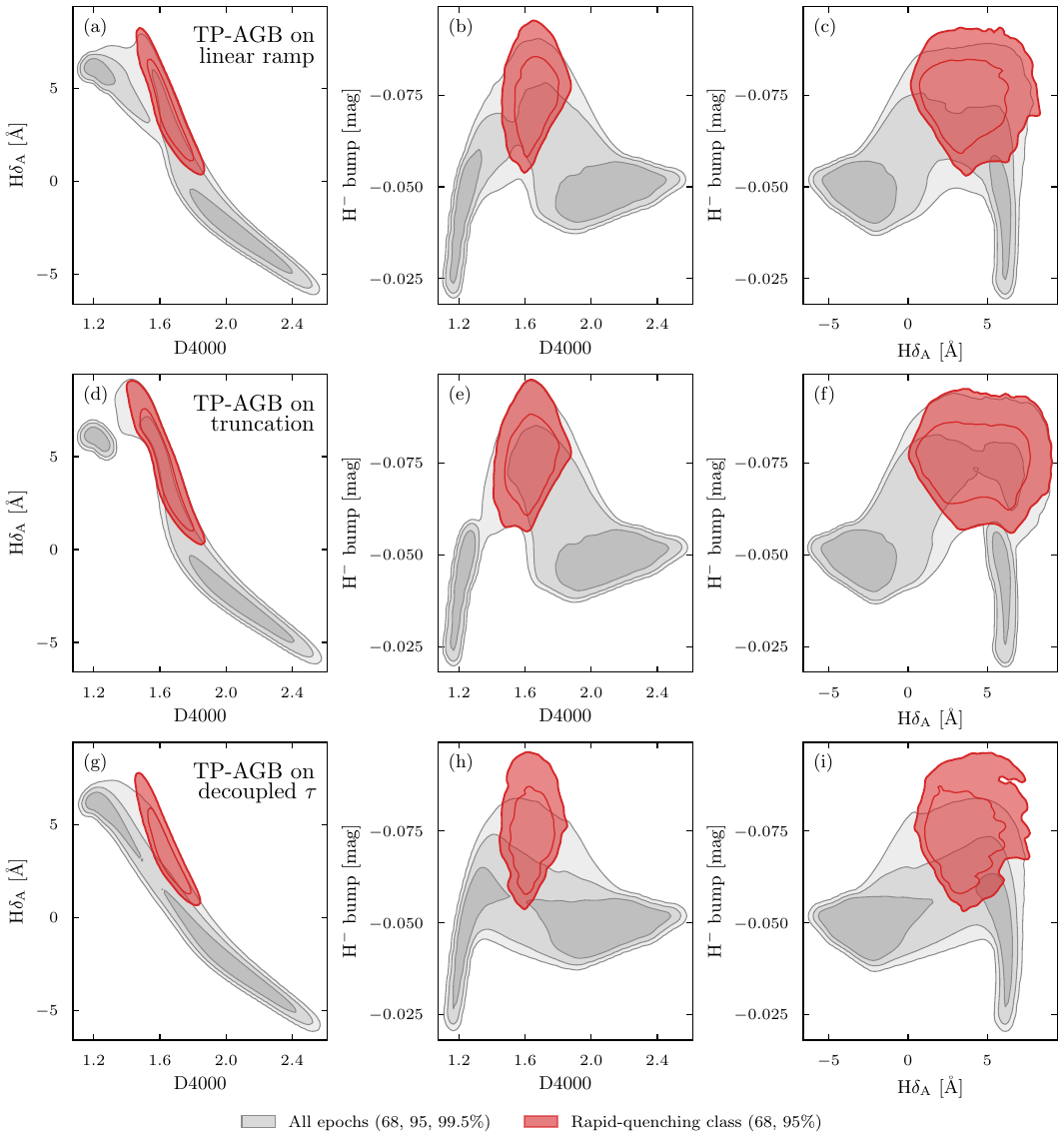}
    \caption{TP-AGB-on index planes for the alternative SFHs described in Appendix~\ref{sec:sfh_families}. Rows show a linear decline, abrupt truncation, and an exponential decline with an independently drawn pre-quenching timescale. Columns and contour fractions follow Figure~\ref{fig:index_planes_agb_on_off}: gray contours describe all retained epochs, and red contours the rapid-quenching class. More negative \hminus{} bump indices indicate stronger features. The detailed distributions and class frequencies vary with SFH, but the rapid-quenching class occupies a similar region with a strong \hminus{} bump in each family.}
    \label{fig:index_planes_sfh_families}
\end{figure*}

    Figure~\ref{fig:index_planes_sfh_families} shows that the rapid-quenching class remains concentrated toward strong \hminus{} bump features in all three families. Its median index is $-0.074$, $-0.076$, and $-0.075$~mag for the linear, truncated, and independently varying-$\tau$ histories, respectively, compared with $-0.075$~mag for the reference exponential family. At a \hminus{} bump uncertainty of 0.010~mag, adding the \hminus{} bump to the optical indices increases completeness by 0.038, 0.026, and 0.052, respectively, and purity by 0.022, 0.010, and 0.036. These are within-family comparisons under the validation procedure of Section~\ref{sec:index}. TP-AGB-off models can yield small gains in some families, but do not show the same combination of completeness and purity improvements.

    The rapid-quenching fractions are 3.58\%, 6.38\%, and 0.84\%, respectively, compared with 1.23\% in the reference ensemble. These fractions vary with the adopted histories and the time spent satisfying the sSFR criteria, which in turn changes the classification baselines. However, they should not be interpreted as observed population abundances. Our tests show that the qualitative \hminus{} bump signature persists across the smooth or abruptly truncated histories considered here. Whether it holds for more general SFHs or a different TP-AGB prescription remains to be tested.

\section{Impact of Spectral Reduction Methods on Index Measurements}
\label{msa-reduction}

The reduction of JWST/NIRSpec MSA spectra can introduce systematic uncertainties in the measured spectral indices, particularly through background subtraction and relative flux calibration. To assess the impact of these effects, we compare the spectral indices measured from two independently reduced sets of spectra for the same 13 galaxies that have both versions. The first uses the DJA global sky subtraction with optimal extraction, while the second adopts the nod2 background subtraction with boxcar extraction from the GTO pipeline (Section~\ref{sec:comparison}). As described in \citet{Scholtz-2026}, the latter 1D spectra are extracted from unrectified 2D frames and corrected for path losses under a point-source assumption. While this assumption is certainly inaccurate for our extended galaxies (as it introduces wavelength-dependent differences in their continuum shapes), it does not strongly affect our local measurements, since the ratio of PSF FWHM between the blue and red ends of the range is of order 20\%.

\begin{figure*}[htbp]
    \centering
    \includegraphics[width=0.94\textwidth]{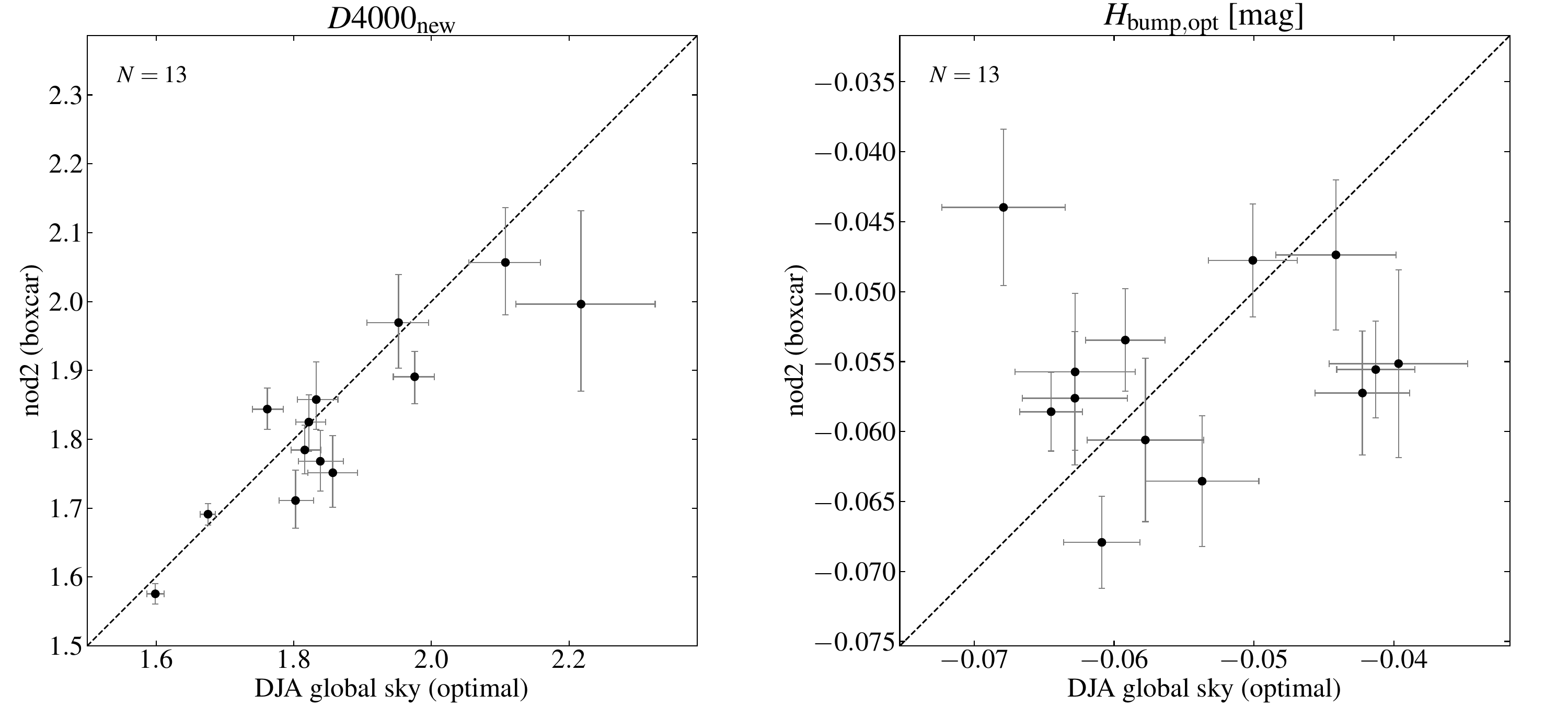}
    \caption{Comparison of $\mathrm{D}4000$ (left) and \hminus{} bump (right) indices measured for 13 galaxies using two reduction methods: DJA global sky subtraction with optimal extraction (horizontal axes) and nod2 subtraction with boxcar extraction (vertical axes). The diagonal line is marked in dashed.}
    \label{fig:compare-reduction-1d}
\end{figure*}

\begin{figure*}[htbp]
    \centering
    \includegraphics[width=0.94\textwidth]{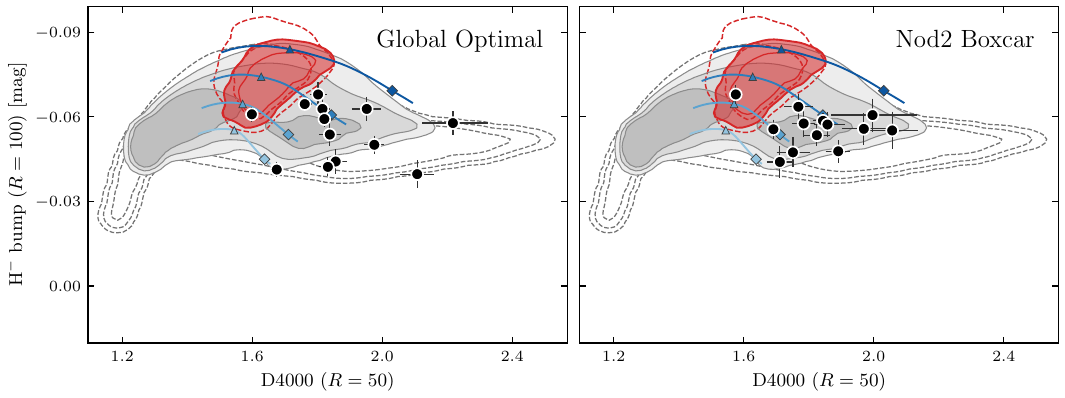}
    \caption{Comparison of spectral indices measured using different reduction methods, following the layout of Figure~\ref{2Dobs}(b). The left panel shows spectra processed with DJA global sky subtraction and optimal extraction, while the right panel shows the same galaxies reduced using nod2 subtraction and boxcar extraction. Although the measured indices exhibit systematic differences, potentially arising from differences in continuum calibration and extraction methods, both are broadly consistent with the TP-AGB-on model.}
    \label{fig:compare-reduction-2d}
\end{figure*}

Figure~\ref{fig:compare-reduction-1d} compares the indices measured using the two reduction methods for 13 galaxies. The D4000 measurements exhibit a strong correlation (Spearman $\rho=0.81$, $p=8.4\times10^{-4}$), with a standard deviation of $0.077$ in the differences between reductions. In contrast, the \hminus{} bump measurements show no significant correlation ($\rho=0.15$, $p=0.63$), with a scatter of $0.011$ mag. However, this is driven entirely by the limited dynamic range of \hminus{} in our sample ($\sim0.03$ mag). After subtracting the typical combined measurement uncertainty of $0.006$ mag in quadrature, the residual scatter is approximately $0.010$ mag, corresponding to a $\sim1\%$ difference in relative flux. Such differences can substantially weaken the correlation between individual measurements. Most importantly, the two measurements scatter about a mean offset of 0.001~mag, much smaller than the offset between the TP-AGB-on and TP-AGB-off models. Hence, all the observed \hminus{} bump indices from both reductions agree with the TP-AGB-on models, supporting the robustness of our conclusion (See Figure~\ref{fig:compare-reduction-2d}).


\bibliography{references}{}
\bibliographystyle{aasjournalv7.1}

\end{document}

%% file: define.tex
\def\apjs{{ApJS}}

\def\h{\hskip -3 mm}

\def\lax{{$\mathrel{\hbox{\rlap{\hbox{\lower4pt\hbox{${\sim}$}}}\hbox{$<$}}}$}}
\def\gax{{$\mathrel{\hbox{\rlap{\hbox{\lower4pt\hbox{${\sim}$}}}\hbox{$>$}}}$}}
\def\simlt{\lower.5ex\hbox{$\; \buildrel < \over {\sim} \;$}}
\def\simgt{\lower.5ex\hbox{$\; \buildrel > \over {\sim} \;$}}

\def\logm10{{$\log (M_{\star,10\ \mathrm{kpc}}/M_{\odot})$}}
\def\logm30{{$\log (M_{\star,30\ \mathrm{kpc}}/M_{\odot})$}}
\def\logm50{{$\log (M_{\star,50\ \mathrm{kpc}}/M_{\odot})$}}
\def\logm100{{$\log (M_{\star,100\ \mathrm{kpc}}/M_{\odot})$}}

\def\mh200b{{$M_{\mathrm{200b}}$}}
\def\mh200c{{$M_{\mathrm{200c}}$}}

\def\mdpl2{\texttt{MDPL2}}

\definecolor{LightGray}{gray}{0.85}
\definecolor{Tab1}{RGB}{114, 158, 206}
\definecolor{Tab2}{RGB}{255, 158,  74}
\definecolor{Tab3}{RGB}{103, 191,  92}
\definecolor{Tab4}{RGB}{174, 199, 232}
\definecolor{Tab5}{RGB}{255, 187, 120}
\definecolor{Tab6}{RGB}{152, 223, 138}
\definecolor{Tab7}{RGB}{255, 152, 150}
\definecolor{Tab8}{RGB}{197, 176, 213}
\definecolor{hpurple}{HTML}{7E16DF}

\newcommand{\temp}[1]{\textcolor{OliveGreen}{#1}} 

\def\hminus{{$\text{H}^{-}$}}